\documentclass{crtbt} % based on article class  and cern-art class
\usepackage{amssymb} 

\DocReference{http://crtbt.grenoble.cnrs.fr/helio/}
\PSLogo{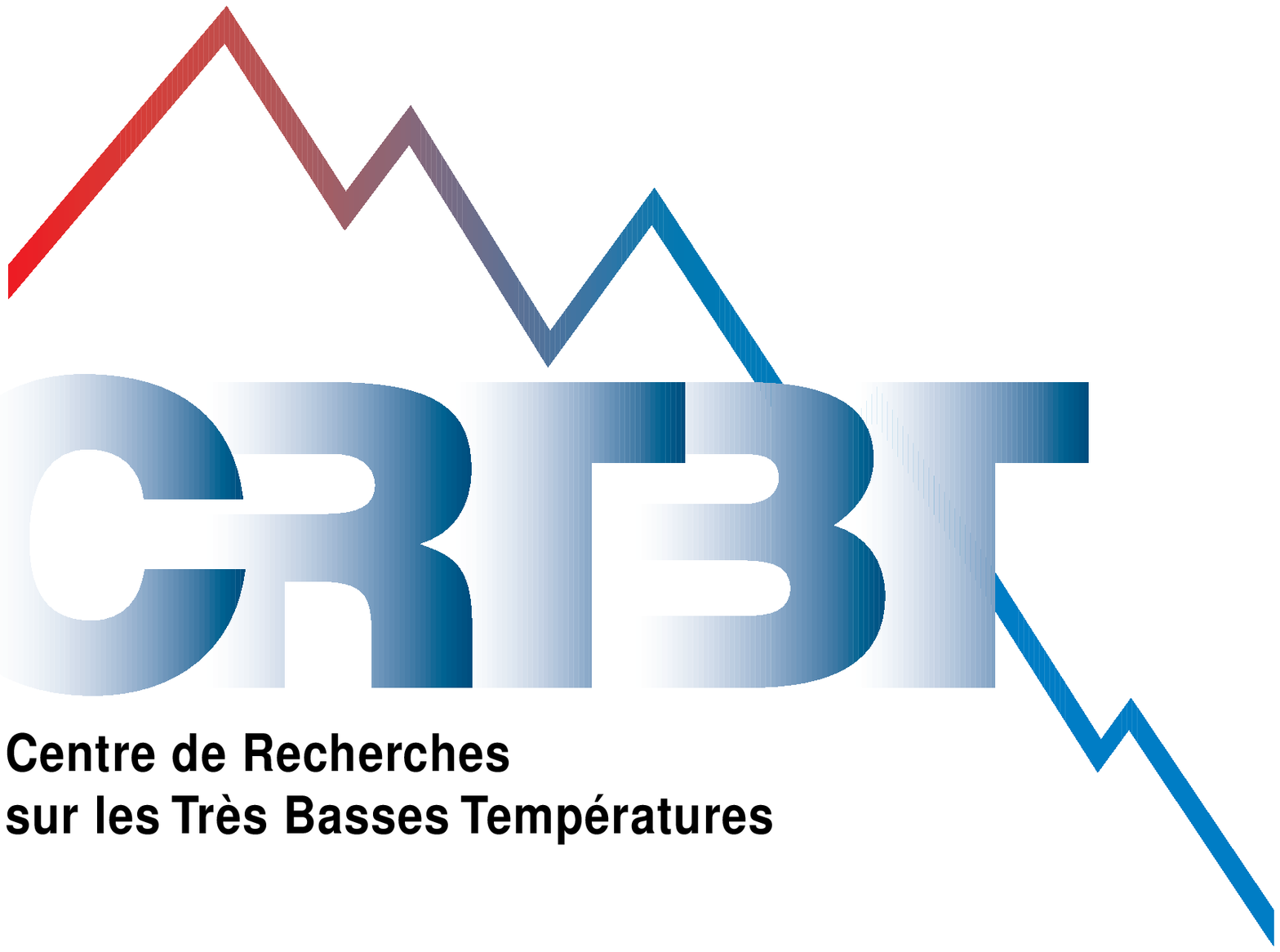}% optional, default is the ...
\LogoHeight{3cm}% optional, default is 2cm 

\Author{P.-E. Roche, B. Castaing$^*$, B. Chabaud and B. H\'ebral} 
\Title{Heat transfer in turbulent Rayleigh-B\'enard\\ convection below the ultimate regime} 

\AdresseBis{$^*$Ecole Normale Sup\'erieure de Lyon\\  46 all\'ee d'Italie, 69364 Lyon Cedex 07, France} % optional 

% optional 
 % optional 
\begin{document} 
\Maketitle 
\Summary{
A Rayleigh-B\'enard cell has been designed to explore the Prandtl (Pr) dependence of turbulent convection in the cross-over range $0.7<Pr<21$ and for the full range of soft and hard turbulences, up to Rayleigh number $Ra\simeq 10^{11}$. The set-up benefits from the favourable characteristics of cryogenic helium-4 in fluid mechanics, in-situ fluid property measurements, and special care on thermometry and calorimetric instrumentation. The cell is cylindrical with $diameter/height=0.5$. The effective heat transfer $Nu(Ra,Pr)$ has been measured with unprecedented accuracy for cryogenic turbulent convection experiments in this range of Rayleigh numbers. Spin-off of this study include improved fits of helium thermodynamics and viscosity properties. Three main results were found. First the $Nu(Ra)$ dependence exhibits a bimodality of the flow with $4-7 \%$ difference in $Nu$ for given $Ra$ and $Pr$. Second, a systematic study of the side-wall influence reveals a measurable effect on the heat transfer. Third, the $Nu(Pr)$ dependence is very small or null : the absolute value of the average logarithmic slope $(dlnNu/dlnPr)_{Ra}$ is smaller than 0.03 in our range of $Pr$, which allows to disciminate between contradictory experiments [Ashkenazi \textit{et al.}, Phys. Rev.Lett. 83:3641 (1999)][Ahlers \textit{et al.}, Phys.Rev.Lett. 86:3320 (2001)].
}% optional, for notes and papers 
%\Object{object of the content}% optional, for minutes and memos 

\section{INTRODUCTION}

Static equilibrium in a column of fluid corresponds to a balance between many parameters such as the weight, the pressure gradient, the temperature difference, etc. The occurrence of a small local perturbation can initiate a global convective motion. Rayleigh-B\'enard convection is a reference configuration for convection studies: a fluid cylinder (height $h$, cross sectional area $S$) located between two horizontal plates is subjected to a temperature difference $\Delta T$ between the plates. In our case the upper plate is regulated at constant temperature and $\Delta T$ results from a constant heat flux $\dot{Q}$ applied at the bottom plate. For high enough $\Delta T$ the convective flow turns into a turbulent regime (for additional references, see for example ref.~\cite{Kadanoff2001}).

For a given fluid at mean temperature T with a mass density $\rho$, and for fixed geometrical conditions, the convective flow is characterized by one single parameter $\Delta T$ or the dimensionless Rayleigh number defined as:

\begin{equation}
Ra=\frac{\alpha g h^3 \Delta T}{ \nu \kappa}
\label{Ra }
\end{equation}

In this expression:\\
- $\alpha$ is the constant-pressure thermal expansion coefficient,\\
- $g$ is the gravity acceleration,\\
- $\nu$ is the kinematic viscosity, $\kappa$ the thermal diffusivity ;  their ratio is the Prandtl number:

\begin{equation}
Pr = \nu / \kappa
\label{Pr}
\end{equation}

The Nusselt number gives the apparent thermal conductivity in the cell: 

\begin{equation}
Nu=\frac{\dot{Q} h}{\lambda S \Delta T}
\label{Nu }
\end{equation}

where $\lambda$ is the fluid thermal conductivity. In a given cell, $Nu$ should depend only of $Ra$ and $Pr$. The influence of the adiabatic gradient on $Ra$ and $Nu$ have been compensated, with the exact correction formula\cite{Tritton1988} :

\begin{equation}
Ra=Ra_{uncorr} \frac{\Delta T - \Delta T_{adiab}}{\Delta T}
\label{Ra_correction }
\end{equation}

\begin{equation}
Nu=Nu_{uncorr} \frac{\Delta T}{\Delta T - \Delta T_{adiab}}.\frac{\dot{Q}- \lambda S \Delta T_{adiab}/h }{\dot{Q}}
\label{Nu_correction }
\end{equation}

An additional correction due to the sidewall conductance is applied, according to the formula proposed in section 4.2.

The unique properties of cryogenic $^4$He allow to control high Rayleigh numbers\cite{Threlfall,Heslot,Wu1992,Chavanne1997,Niemela2000}. In particular, these were used in Grenoble to reach $Ra$ higher than $2.10^{12}$  in ``high'' cells (aspect ratio 1/2, h = 20 cm) : in such conditions they observed for the first time the Kraichnan regime\cite{Chavanne1997} (also called the \textit{ultimate regime}) and its asymptotic limit\cite{RochePRE2001}, both predicted forty years ago\cite{Kraichnan1962}. 

Helium gives also the opportunity to easily vary the Prandtl number\cite{Ahlers1978} over an unusual range. We have done a specific study of the $Pr$ variation effect in a small size cell (called the mini-cell, aspect ratio $1/2$, h = 2 cm), for $3.10^6 < Ra < 10^{11}$ corresponding to the soft and hard turbulence regimes where experimentally controlled $Pr$ can be achieved independently of $Ra$ (the lowest explored $Ra$ in this work is below $10^4$ and the convection threshold was found to be the same as in the large cells, around $4.10^4$). This $Ra$ range fully covers the turbulent regimes preceding the transition towards the ultimate regime and allows a comparison with other experiments using various fluids, and also several models. 

\begin{figure}
\centerline{\includegraphics[height=2.5in]{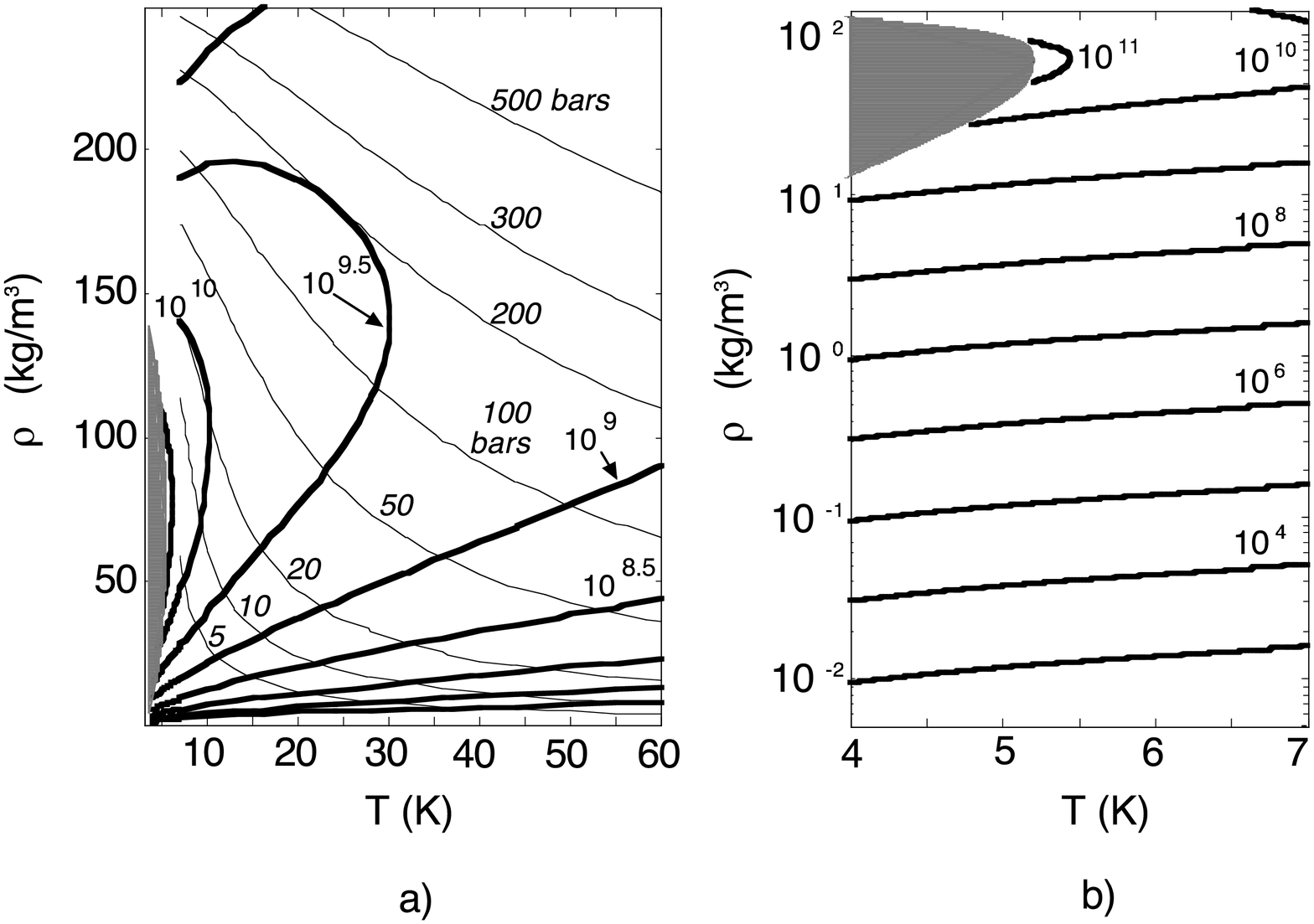}}
\caption{a-Isobars and lines of maximum attainable $Ra$ in the density-temperature plane. b- Semi log enlargement of the plot a, corresponding to the experiments described in the text. The maximum $Ra$ achieved within Boussinesq conditions in the $2\,cm$ high cell is $10^{11}$. The dark area corresponds to the 2-phases region.} 
\label{fig:RaMaxi}
\end{figure}

Lines of maximum attainable Rayleigh numbers, hereafter called iso-$Ra_{max}$ curves, have been calculated at various pressures and temperatures for the mini-cell and for the maximum temperature difference $\Delta T_{max}$ compatible with a Boussinesq criterion\cite{Tritton1988}, here defined as: $\alpha \Delta T < 20\%$. This criterion is close to the one respected in the experimental data for which $\alpha \Delta T < 21\%$. In density-temperature coordinates iso-$Ra_{max}$ curves are shown on figures~\ref{fig:RaMaxi}-a and \ref{fig:RaMaxi}-b  as the thick full lines. Close to the critical point, the divergence of $\frac{1}{\nu \kappa}$, easily accessible in $^4$He ($2.2\ bars$, $5.2\ K$) is clearly illustrated by the extremely high $Ra$ which can be obtained in reasonable experimental conditions ($\Delta T > 1\ mK$). The thin lines on figure~\ref{fig:RaMaxi}-a correspond to isobars.

In the large cells\cite{Chavanne1997,RocheEPJB2001}, the $Pr$ variation is only obtained for $Ra$ above $10^{10}$. Clear understanding of the $Pr$ variation is difficult in these cells due to the occurrence of the ultimate regime. With cell dimensions divided by ten, the $Pr$ variation is already observable around $Ra = 10^7$.

It is worth noticing that the divergence in $Ra_{max}$ is due to $C_p$ in the relation $\frac{1}{\nu \kappa}=\frac{C_p \rho ^2}{\eta \lambda}$, $\eta$ being the fluid viscosity and $C_p$ the specific heat at constant pressure. The Prandtl number varies as: $Pr=\frac{C_p \eta}{\lambda}$ . The $C_p$ divergence appears far away from the critical point and gives a long range effect to the rapid variation of $Ra$ and $Pr$. In various experimental conditions, $\Delta T$ has been widely varied up to three decades at roughly constant temperature and density. This is the largest excursion ever achieved in Rayleigh-B\'enard experiments within Boussinesq condition. This gives access to power law exponent $Nu$ versus $Ra$ at constant $Pr$ and independently of the fluid properties knowledge. Indeed, as temperature and density in the bulk of the flow are almost constant, the same hold for the He properties\footnote{In our measurements, the average temperature and density slightly differ from one point to another. These variations in the experimental conditions are precisely measured and the fluid properties are recalculated for each point. If we make the unrealistic hypothesis that the fluid properties variation is estimated with a $100\%$ error, the resulting uncertainty of the effective power law of the Nu(Ra) dependence would be less than $4\%$.}. Thus, the power law exponent of $Nu(Ra)$ is independent of possible error on the fluid properties, if we except the adiabatic gradient correction. However these properties have to be known precisely for the $Nu$ versus $Pr$ dependence studies.

When this experiment was designed, the situation was the following : two experiments\cite{Ashkenazi,Ahlers2001} were given contradictory results, each being is agreement with a different theory. The first one, conducted over nearly 2 decades of $Pr$ ($1< Pr < 93$) found a $-0.2$ exponent for the effective $Nu(Pr)$ power law, while the second experiment's data can be fitted with a -0.01 exponent over 0.9 decade of $Pr$ ($4< Pr < 34$). Our aim was to elucidate the controversy and to expand the explored $Pr$ range below $Pr=1$. Our experiment allows to vary the Prandtl over 1.5 decade, that is $0.7 < Pr < 21$ and for a large excursion of $Ra$ numbers. The references~\cite{Liu1997,Verzicco1999,Kerr2000,Xia2002} present $Pr$-dependence studies conducted in the Rayleigh-B\'enard geometry for lower $Ra$, much lower or higher $Pr$, or for a much smaller $Pr$ range. The thermal control and measurement accuracy of our experiment are unprecedented in cryogenics convection experiments for $Ra>3.10^6$. It revealed two unexpected effects : the side-wall effect and the bimodality.

This paper gives a detailed description of the apparatus (section 2) and improvements of $^4He$ properties fits (section 3). In section 4, we remind the three main results : the bimodality of the flow, the side-wall effect and the Prandtl number dependence. Section 5 proposes some perspectives for convection studies.

\section{INSTRUMENTATION}

\subsection{The Rayleigh-B\'enard cell design}
\begin{figure}
\centerline{\includegraphics[height=2.5in]{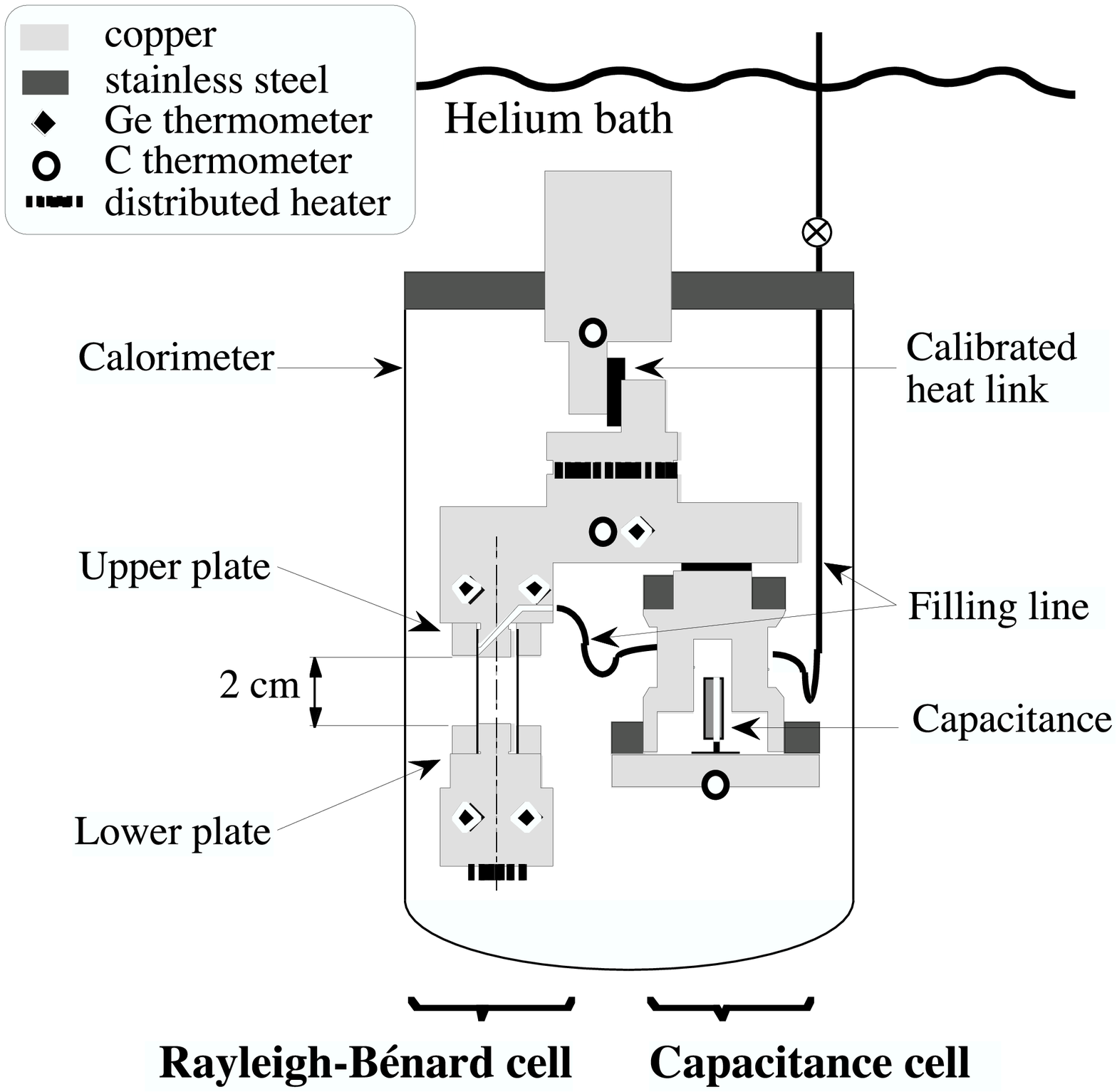}}
\caption{Scheme of the experimental set-up.} 
\label{fig:set_up}
\end{figure}

\begin{figure}
\centerline{\includegraphics[height=2.5in]{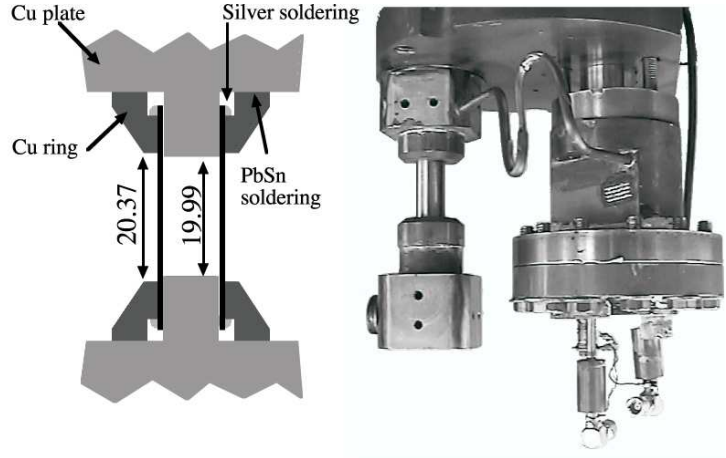}}
\caption{4 a- The Rayleigh-B\'enard cell. 4b- Photograph of the experimental set up (same arrangement as on fig.~\ref{fig:set_up} ).} 
\label{fig:paroi_photo}
\end{figure}

The experimental set-up, presented on figure~\ref{fig:set_up}, is placed in a cryogenic vacuum. The $1/2$ aspect ratio mini-cell ($h = 2\,cm$), is also shown on figure~\ref{fig:paroi_photo}. The stainless-steel cylindrical wall is $0.25$ mm thick. It can hold pressures up to several tens of bars. The upper plate is part of a main Cu flange which ensures the thermal link to the liquid He bath, through a brass plate of measured thermal resistance (53 K/W at 4.3K) and a high conductivity Cu post\cite{Chavanne1996,Chavanne1998,ChavanneTHESE,RocheTHESE}. The brass plate acts as a thermal resistance which allows to regulate the top plate at temperatures different from that of the He bath. The lower plate is also made of copper. In such a helium/copper set-up and for the $Ra$ numbers explored in this study, the plates properties (finite conductivity and heat capacity) do not alter the dynamical formation of the coherent structures (plumes,...)\cite{ChaumatETC,Verzicco_plate}.

Special care has been taken during the cell assembly. The roughness of the plates surfaces in contact with the fluid is estimated less than 2 $\mu$m. The silver soldering of the wall and the copper rings ensure the connection to the plates. For all the cells developed in our group, the side wall design is chosen in order to have a perfect cylinder along all the active length of the cell\cite{Chavanne1996,Chavanne1998,ChavanneTHESE,RocheTHESE}. The plates parallelism is guaranteed by special machining procedure performed \textit{after} the silver soldering: the distance between plates is 19.99 $\pm$0.02 mm. 

The cell is filled through a capillary closed with a cold needle valve located in the main helium bath. This capillary is connected in series with a capacitance cell, that will be further described, and the Rayleigh-B\'enard cell. 

\subsection{Experimental procedure}

Each cell plate holds two Ge resistors from the same batch with close resistance values and temperature dependence. The upper flange is temperature regulated with a PID (Proportional-Integral-Derivative) analogue regulator with a fifth Ge resistor. The temperature difference $\Delta T$ between the plates in the Rayleigh-B\'enard cell is determined from the measurement of the ratio of two Ge resistors using a resistance ratio bridge operating at 30 Hz with 1 $\mu$A current amplitude\cite{castaing1982}. The resistance ratio variation with and without heating gives $\Delta T$ through the calibration of the resistors and the additional measurement of the upper plate temperature. This procedure is valid even for $\Delta T$ larger than $1\ K$, as checked with the direct temperature measurement of each plate. The ratio without heating (Òzero $\Delta T$Ó) is monitored during twelve hours before and after each measurement cycle. In less than one hour the equilibrium value is obtained, except for the data close to the critical point: in such conditions the thermal diffusion time diverges and the used zero $\Delta T$ is the one obtained at a lower density. 

A ratio variation of $10^{-5}$ at $5\ K$ corresponds to $25\ \mu K$ for $\Delta T$. The stability of the set-up and electronic apparatus is better than $30\ \mu K$ over $12\ hours$. The radiation heat losses are estimated to be around 10 nW which gives typically $50\ \mu K$ for $\Delta T$. This is of the order of the adiabatic gradient temperature difference\cite{Tritton1988,ChavanneTHESE} in the $2\,cm$ high cell, this effect limiting the smallest achievable $\Delta T$. More than 3 decades of variation of $\Delta T$ have been achieved for given mean temperature and density, from below the convection onset up to the turbulent regimes. The Boussinesq criterion is $\alpha .\Delta T<21\%$. Besides the conductivity and viscosity never vary by more than $8\%$ between the bottom and top of the cell. Still between the bottom and top of the cell, the constant-pressure heat capacity and the expansion coefficient $\alpha$ vary by less than $10\%$ for $88\%$ of the points, they vary by less than $20\%$ for $95\%$ of the points and by less than $40\%$ for all the points.  In large cells, an original thermocouple technique is more appropriate than a resistance bridge to measure the temperature difference: the $\Delta T$ zeroing procedure is not compatible with the large thermal relaxation times.

A four wires voltage standard (Electronics Development Corporation) provides a constant heating power on the lower plate. In order to limit local overheating, the heater is distributed on the surface. All the copper pieces are made out of a commercial non-annealed Cu, which was characterized in another experiment: its thermal conductivity is around $400\ W/m.K$. We have measured the wall thermal conductance between 4.5 and $6\ K$. For cross-validation, two measurements have been done with an empty cell and with helium at $80\ g/m^3$. After subtraction of the diffusive helium contribution both results agree within $2.5\%$: in this difference $2\%$ are explained by the cell design\cite{RocheTHESE}. This side-wall contribution is described by the fit $-40.1 + 44.75.T$ in $\mu W/K$ including the conductance of the copper heating wires ($33\ \mu W/K$ at $4.5\ K$ with $20\%$ uncertainty) going to the lower plate. The lower plate heat capacity, as measured by a relaxation method, is $73\ mJ/K$ including addenda (Ge resistor holders and copper ring). The brass plate heat leak has a measured resistance of $53\ K/W$ at $4.3\ K$. 

\subsection{Densitometry}

In order to determine the $Ra$ and $Nu$ with an absolute resolution of a few percent, a density accuracy of $1\%$, at least, is needed. The dead volume coming from the filling capillary going to room temperature is too large to determine with enough precision the He density during the cell filling procedure. We thus have performed an in-situ measurement using a capacitive probe located in a specific cell, in order to have a much better resolution. The density is extracted from the Clausius-Mosotti relation: $\frac{\epsilon -1}{\epsilon +2}=\frac{4\pi \zeta}{3M}\rho$, where $\epsilon$, $\rho$, $M$, $\zeta$ are respectively the permittivity, the density, the molar mass and the polarisability of helium ($\zeta = 0.123296$ $cm^3/mol$)\cite{Donnelly1998}. 

Two capacitances are placed in a ratio bridge. The two porous frames of the ``active'' capacitance ($C\simeq17.5$ pF) are made out of printed circuits and $0.1\,mm$ separated. This capacitance is located in the capacitive cell and totally immersed in helium. Under these conditions the mechanical dependence with pressure effects is minimized. On the inner part of each frame a circular electrode (16 mm diameter) is engraved together with a guard ring. Special attention in the design reduces differential contraction effect and parasitic capacitances : no spurious effect were   detected and no temperature effects were observed. The other capacitance ($C_r = 9.4\,pF$) made out of mica, is located at $4.2\ K$ in the calorimeter vacuum. It is the reference one. 

The bridge operates at 3 kHz. The ratio between both capacitances is a direct measurement of $\epsilon$. The density measurement range is $0-140\ kg/m^3$, under pressures from 0 up to 7 bars and temperatures between 4.5 and $6.5\ K$. Over ten days the stability is $10^{-5}$ ($\pm 40\,g/m^3$), that is better than $0.1\%$ in density for $Ra$ above $2.10^6$. Two calibrations of the capacitance ratio at the beginning and at the end of the experiment agree within $30\,g/m^3$. The signal averaged over $30~s$ has a resolution of $10^{-7}$ (less than $1\,g/m^3$), which can be maintained over a few hours.

In principle $k=\frac{C}{\epsilon C_r}$ should be a constant over the density range. We have achieved low and high densities calibrations. During the low pressure calibration, the cell, connected to a few litres reservoir at room temperature, is regulated at $5.432\ K$. For pressures lower than 1000 mbar, no condensation occurs in the filling line. In order to have stable operation conditions we restrain the low-pressure calibration below 350 mbars: a precise pressure measurement gives access to the density through ref.~\cite{McCarty1990} . At high density we measured the density following a procedure described below for the absolute temperature calibration (see ÒThermometryÓ). For example, with 31 $\mu$W heating power applied on the bottom plate, we extract a density of $112.73\ kg/m^3$ from the measured boiling temperature ($4.7088\ K$ with $\Delta T = 6.5 mK$).

\begin{figure}
\centerline{\includegraphics[height=2.5in]{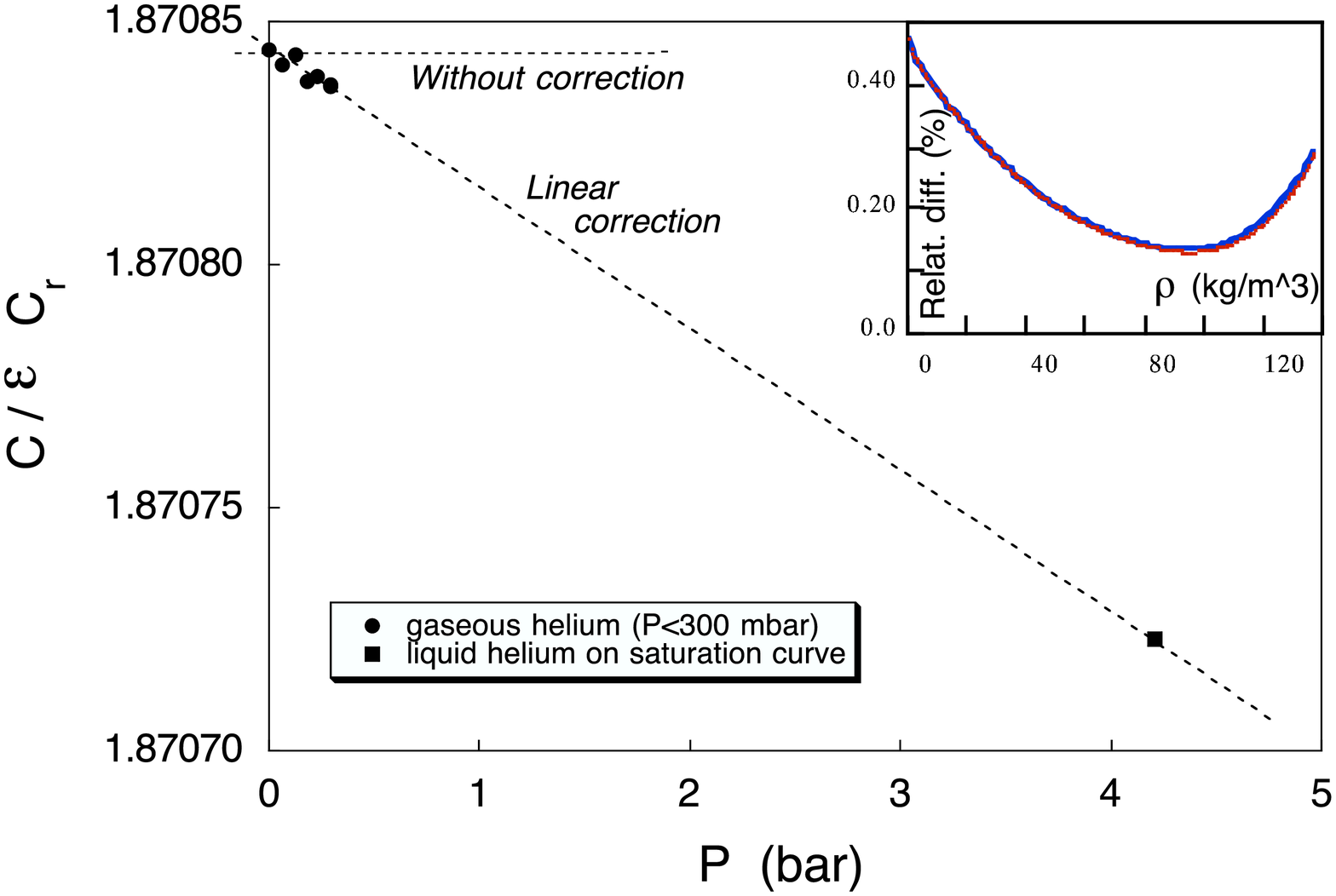}}
\caption{Calibration of the density measurement cell. Insert: Difference, in percent, between the density estimations with and without pressure correction.} 
\label{fig:etal_capa}
\end{figure}

The calibration results are summarized on figure~\ref{fig:etal_capa}  where $\frac{C}{\epsilon C_r}$ is plotted as a function of the pressure $P$. $\frac{C}{\epsilon C_r}$ varies by $6.10^{-5}$ over the whole range and we have assumed a linear variation versus $P$. Such behaviour is typical of a residual mechanical deformation. On the figure insert the difference between density with and without the linear correction is plotted versus $\rho$. This $\delta \rho$ is less than $0.5\%$ and goes through a minimum at 0.15$\%$ close to the critical density. In all the following we obtain the density from the linear pressure correction and we estimate the $\rho$ uncertainty to be about $0.1\%$.

\begin{figure}
\centerline{\includegraphics[height=2.5in]{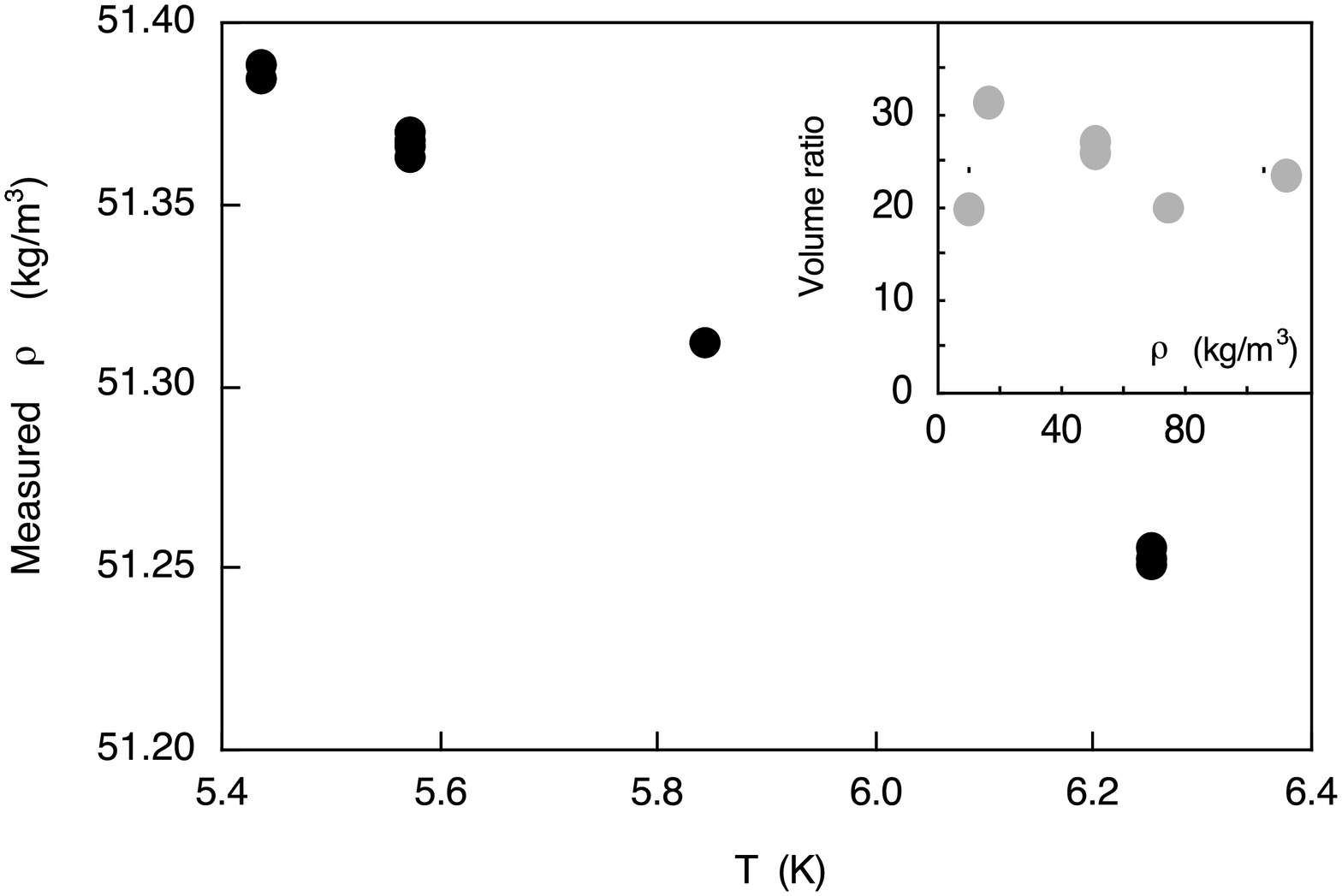}}
\caption{Measurement of the helium compression effect in the filling capillary. Insert : Measured ratio of the cells volume divided by the $4.22\ K$ capillary volume versus density.} 
\label{fig:capilaire}
\end{figure}

When both cells are filled and  the needle valve is closed, no temperature dependence of the capacitive signal is expected. However a $0.3\%$ tiny reproducible (on a few months scale) variation was observed as seen on figure~\ref{fig:capilaire}. This was explained by the helium compression in the upper part of the filling line close to the needle valve in thermal contact with the main helium bath at $4.22\ K$. We have evidenced a linear correlation through the comparison of the total measured density and the calculated density in the capillary. The slope, plotted on the insert of figure~\ref{fig:capilaire} is the ratio between the total volume and the capillary dead volume : the value 23 $\pm$4 is in fair agreement with a less precise value extracted from a geometric determination. The $\pm 4$ uncertainty on the volume ratio (due to the scatter of points) corresponds to a measured density uncertainty less than $0.06\%$ on the whole densities and temperature range. It confirms by an independent way our formerly quoted $0.1\%$ density uncertainty. Note that this capillary effect has strictly no influence on the density measurement in the cells.

\subsection{Thermometry}

\begin{figure}
\centerline{\includegraphics[height=2.5in]{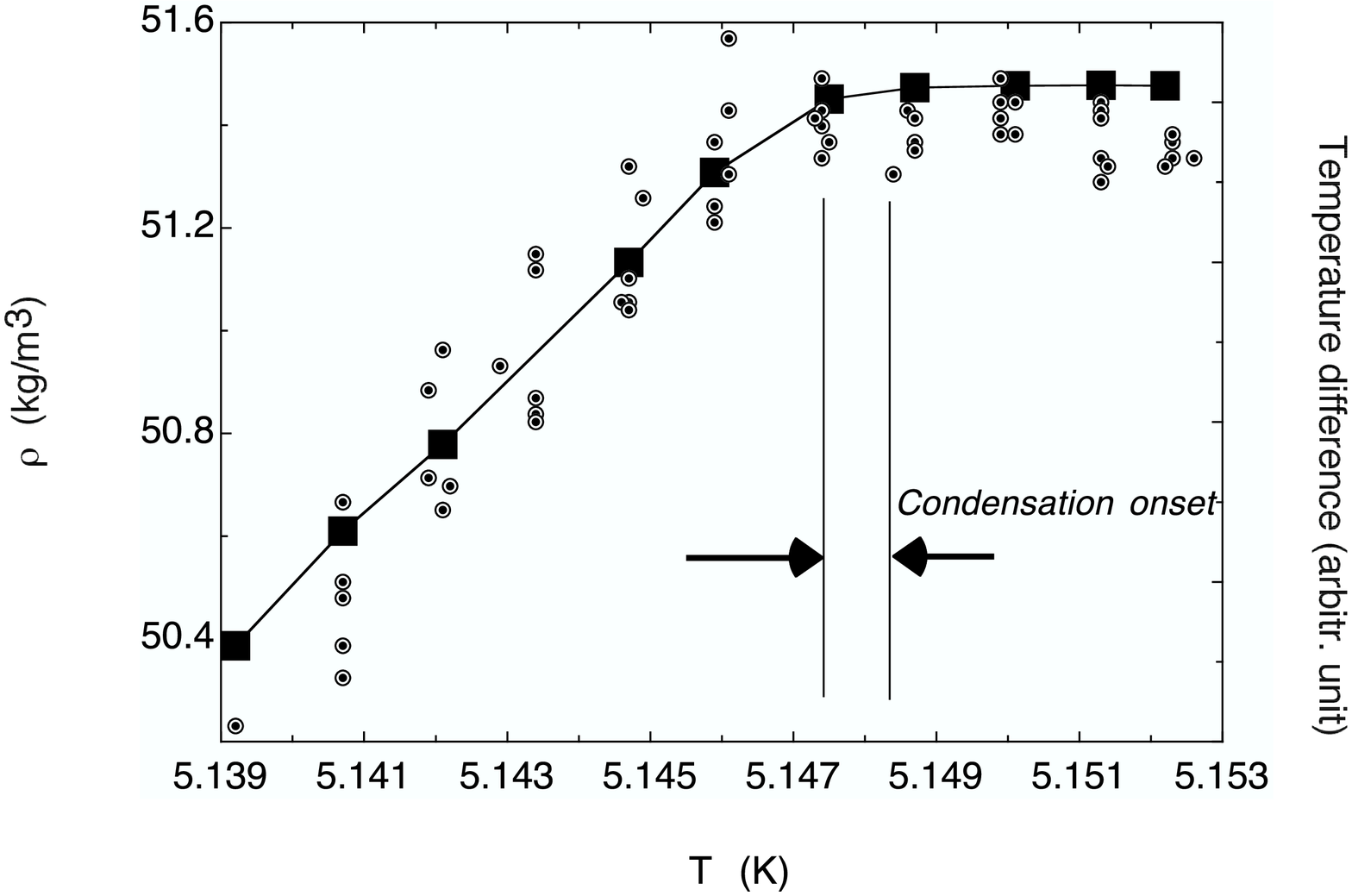}}
\caption{Full squares: Cell density versus top plate temperature around the condensation onset. Dotted circles: PlatesÕ temperature difference in arbitrary units.} 
\label{fig:condensation}
\end{figure}

A one millikelvin uncertainty on the mean temperature gives an uncertainty up to $1\%$ on the $Ra$ and $Nu$ values, in the range of temperature and density of this experiment. In order to compare the various data issued from several references\cite{Donnelly1998,McCarty1990,Moldover1969,Kierstead1971,Kierstead1973,McCarty1973,Acton1977,Acton1981,Agosta1987}, we use the ITS-90\cite{ITS1990} critical temperature: $T_c = 5.1954\ K$ and adjust the thermometer in-situ calibration onto that value. This calibration procedure is illustrated on figure~\ref{fig:condensation} at a density of $51.5\ kg/m^3$ where the condensation is expected at: $T = Tc-47.4\ mK$. We apply a small 125 $\mu$W heating power and monitor both $\Delta T$ across the Rayleigh-B\'enard cell and the gas density $\rho _{gas}$ in the capacitive cell. The temperature of the upper plate is slowly lowered. The sharp drop of $\Delta T$ and $\rho _{gas}$ is the signature of the condensation in the cell. It is worth to note that the density measurement is more precise than the $\Delta T$ one\cite{Ashkenazi,Chavanne1998} to identify condensation and enables a $1\ mK$ resolution. 

\section {$^4$He PROPERTIES}

\subsection{Thermal expansion coefficient}

\begin{figure}
\centerline{\includegraphics[height=2.5in]{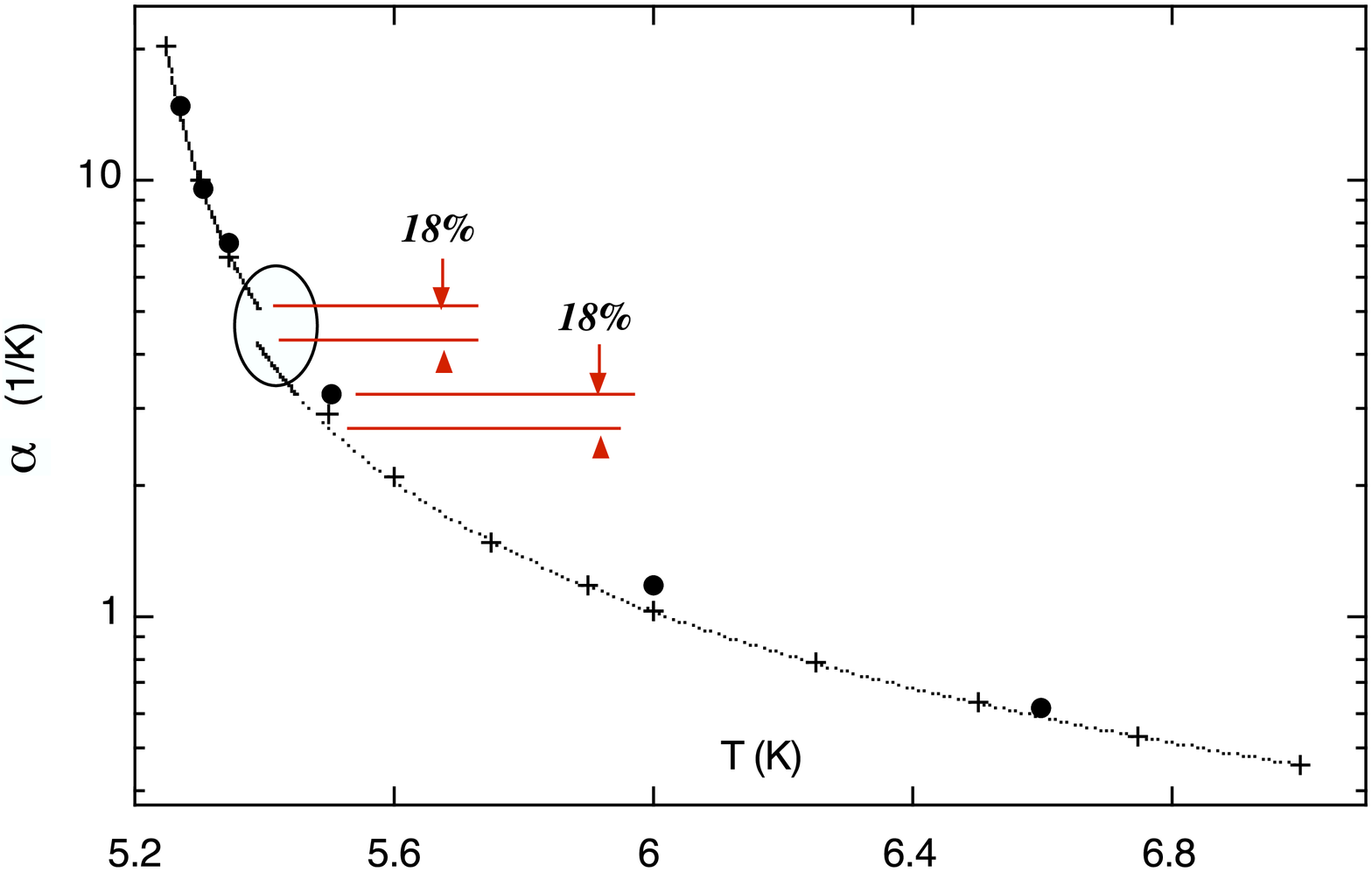}}
\caption{Comparison between the thermal expansion coefficients at density $\rho =74.0 kg/m^3$ ($\pm 0.6 kg/m^3$ for the experimental data). Present measurements are in full circles. Kierstead\cite{Kierstead1973} ($T<5.4 K$) and McCarty and Arp\cite{McCarty1990} ($T>5.4 K$) fits (lines) have been extrapolated in the circled area. The XHePak fit (pluses) is the commercial fitting package cited in ref.~\cite{Niemela2000}}
\label{fig:raccordement}
\end{figure}

The reference fits of Arp and McCarty\cite{McCarty1973,McCarty1990} account for $^4$He thermodynamics properties over a wide temperature and pressure range ($0.8-15000\ K$, $0-2000\ MPa$) but ignores the critical divergence, which has been fitted by Kierstead\cite{Kierstead1971,Kierstead1973}. Unfortunately, the temperature and density validity ranges of the two fits do not overlap. Besides, extrapolation of the thermal expansion coefficient from both fits suggests up to 25$\%$ discrepancy between these fits, which is incompatible with the extrapolation uncertainty. Figure~\ref{fig:raccordement} illustrates this discrepancy for a density of $\rho = 74\, kg/m^3$.

\begin{figure}
\centerline{\includegraphics[height=2.5in]{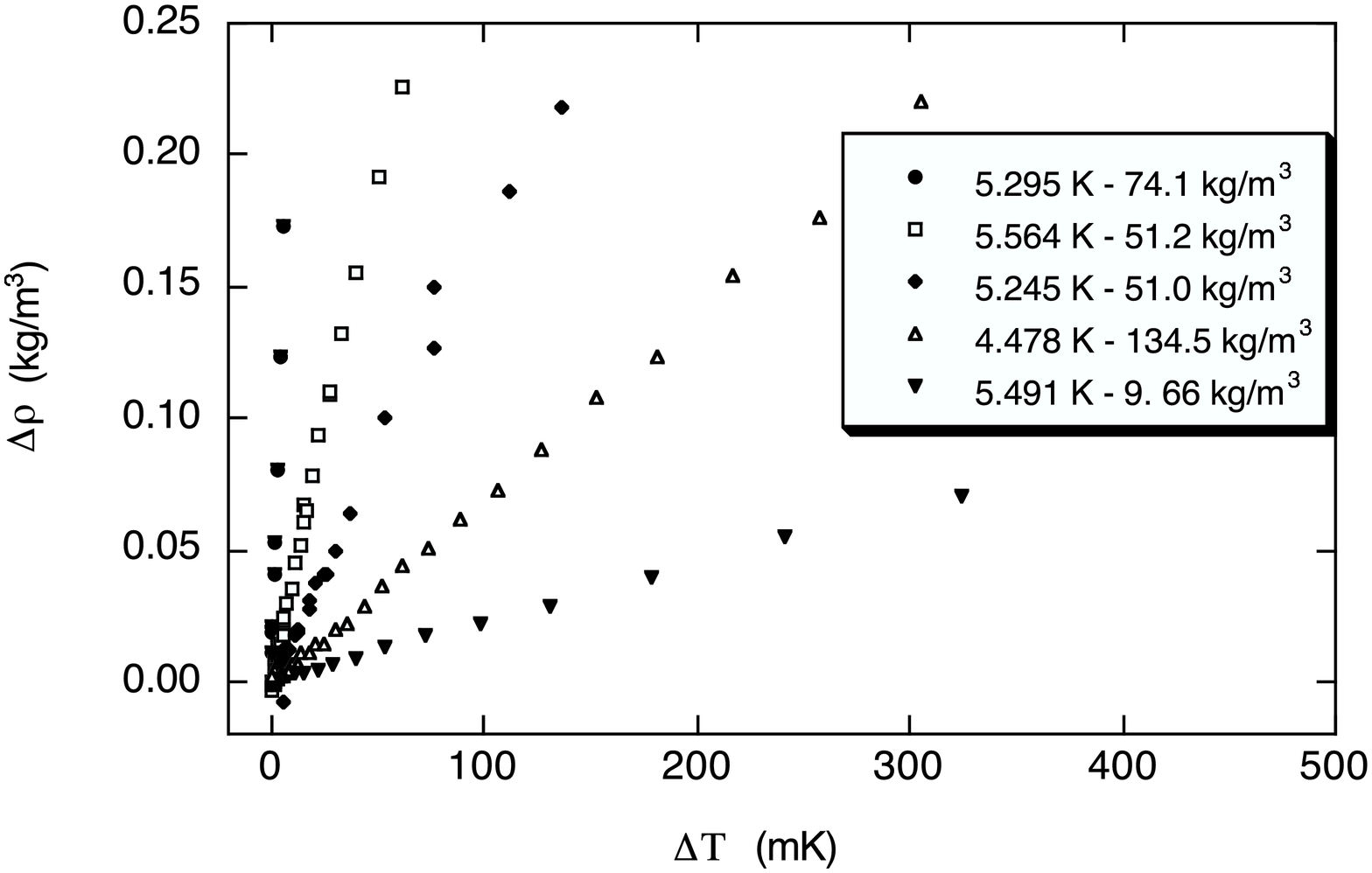}}
\caption{Capacitive measurement of the $\rho$ increase in the capacitive cell due to the Rayleigh-B\'enard cell heating versus the platesÕ temperature difference in the convection cell.} 
\label{fig:alpha}
\end{figure}

In our experiment, the high sensitivity of the capacitive cell gives access to $\alpha$, in a temperature and pressure range overlapping both fits. The Rayleigh-B\'enard cell heating increases the density in the capacitive cell. For low $\Delta T$ across the convection cell, the density variation is given by: $\delta \rho= \frac{\Delta T}{2} \rho \alpha \frac{v_2}{v_1+v_2}$ , where $v_1$ and $v_2$ are the capacitive cell and Rayleigh-B\'enard cell volumes respectively. The results are shown on figure~\ref{fig:alpha}. We extract the slope of these curves for $\Delta T$ going to 0, for given temperature and density conditions. We have also done a correction\cite{RocheTHESE} to take into account the volume of the $4.2\ K$ filling capillary below the needle valve: this correction represents a few percent at low and high densities and is smaller than $0.5\%$ between 40 and $80\ kg/m^3$. Each point corresponds to a Temperature-Density condition which fully falls into the validity range of either one of the two fits.

\begin{figure}
\centerline{\includegraphics[height=2.5in]{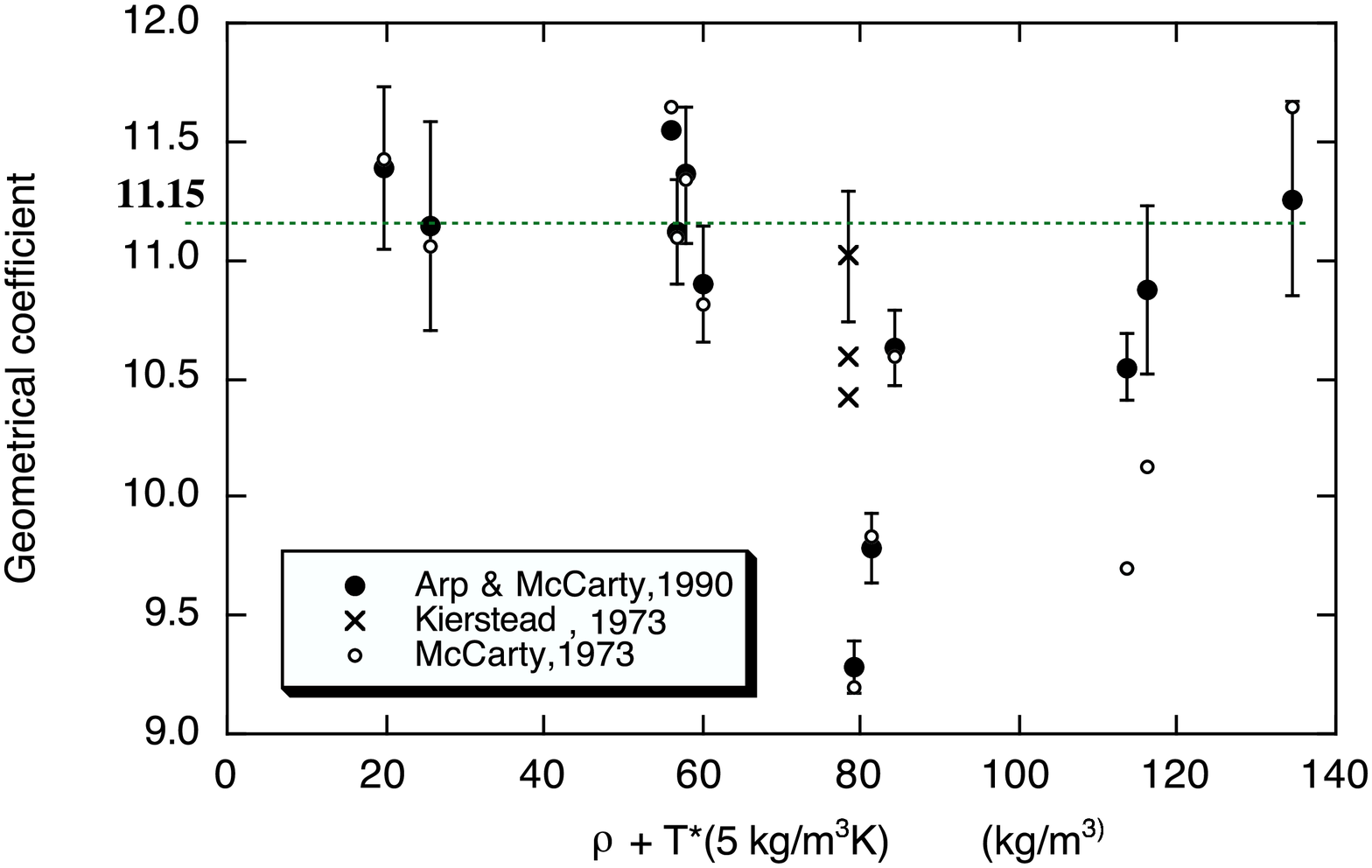}}
\caption{Geometrical coefficient (see text) versus $\rho +5T$ for the determination of the helium expansion coefficient.} 
\label{fig:coefficient}
\end{figure}

Determination of the geometrical coefficient $\frac{v_2}{v_1+v_2}$ is illustrated on figure~\ref{fig:coefficient} (the capillary volume introduces less than $1\%$ correction on this formula and this correction is not shown here but it is taken into account in the analysis of ref.~\cite{RocheTHESE}).  Figure~\ref{fig:coefficient} shows  the quantity $\alpha _{est} / {\frac{\alpha v_2}{v_1+v_2}}=\alpha _{est} / {\frac{2 \delta \rho}{\rho \Delta T}}$ where $\alpha _{est}$ is an estimated $\alpha$ value from the literature\cite{McCarty1990,Kierstead1973,McCarty1973}. The x-axis is chosen in order to avoid the degeneracy for the data taken at same density but  at various temperatures. Most of the data are compatible with the value 11.15 $\pm0.25$.

As expected the data using the values from ref.~\cite{McCarty1990} are very reliable far from the critical point, but they need a correction which rises up to about $20\%$ when approaching this regime. The data from ref.~\cite{McCarty1973} are in good agreement with those of ref.~\cite{McCarty1990} at low density but need a correction of several percent at high density. In the critical region ref.~\cite{Kierstead1973} appears to give the best agreement, as illustrated on Figure~\ref{fig:raccordement}. The various $\alpha$ values are summarized in Table 1.

\begin{center}
\begin{tabular}{|c|c|c|c|c|c|c|}
\hline
% after \\ : \hline or \cline{col1-col2} \cline{col3-col4} ...
$\rho $ & $T$ & $\alpha_{measur.}$ & $\frac{\Delta \alpha_{measur.}}{ \alpha_{measur.}} $ & $\alpha_{MCA} $ & $\alpha_{Kierstead} $ & Measur. vs.\\
& & & & & & Fit disagr.\\
$(kg/m^3)$ & $(K)$ & $(K^{-1})$ & $(\%)$ & $(K^{-1})$ & $(K^{-1})$ & $(\%)$\\
\hline
16.76 & 5.114 & 0.387 & 5 & 0.395 &  & -2 \\
\hline
16.70 & 6.270 & 0.266 & 6 & 0.266 &  & 0.1\\
\hline
51.4 & 5.438 & 2.23 & 3 & 2.311 &  & -3.5\\
\hline
51.4 & 5.573 & 1.72 & 4 & 1.716 &  & 0.3\\
\hline
51.3 & 5.844 & 1.11 & 4.5 & 1.128 &  & -2\\
\hline
51.25 & 6.252 & 0.762 & 4.5 & 0.745 &  & 2.5\\
\hline
74.60 & 5.268 & 14.7 &  & & 14.001 & 5\\
\hline
74.31 & 5.305 & 9.48 & 4.5 &  & 9.3682 & 1\\
\hline
74.21 & 5.346 & 7.18 & 2.5 &  & 6.7188 & 6.5\\
\hline
74.06 & 5.503 & 3.21 & 3 & 2.671 &  & 18\\
\hline
73.87 & 5.999 & 1.170 & 3.5 & 1.027 &  & 13\\
\hline
73.73 & 6.600 & 0.616 & 3.5 & 0.587 &  & 5\\
\hline
112.3 & 5.260 & 0.349 & 5.5 & 0.340 &  & 2.5\\
\hline
112.6 & 4.725 & 0.478 & 3.5 & 0.453 &  & 5.5\\
\hline
134.6 & 4.490 & 0.1104 & 5.5 & 0.1115 &  & -1\\
\hline
\end{tabular}
\noindent {\bf Table 1}~: $\alpha_{measur.}$ : expansion coefficient measurements, $\Delta \alpha_{measur.}$ total uncertainty on $\alpha_{measur.}$, $\alpha_{MCA} $ :  expansion coefficient estimated with McCarty and Arp 1990 fit\cite{McCarty1990}  ; $\alpha_{Kierstead}$ :  expansion coefficient estimated with Kierstead fit\cite{Kierstead1973}.
\end{center}

In our Rayleigh-B\'enard data analysis, the $\alpha$ coefficient and other thermodynamical coefficients such as $(C_p-C_v)$ are obtained from the first order partial derivatives of the pressure versus temperature or density. Such a way of deriving properties ensures the self-consistency between thermodynamics parameters. As a consequence, taking into account our measurements we have corrected the fits at their source, that is directly on the fit of $(\partial P / \partial \rho)_{T,MCA}$ of ref.~\cite{McCarty1990} : \\
\\
$(\partial P / \partial \rho)_T=(\partial P / \partial \rho)_{T,MCA} [ 1+ F_T.F_\rho ]$
\\
\\
with 
$F_T=4.62-0.658T$ and
$F_\rho =0.246-0.00117(\rho-67)^2$
\\
\\
Thus our recommendations are the following : in the critical zone as defined by Kierstead ($55.7<\rho <83.5~kg/m^3$ and $T<5.362~K$) use the Kierstead fit, out of this range, use the Arp and MacCarty\cite{McCarty1990} fit with the above correction if both $F_T$ and $F_\rho$ are positive and if the total relative correction $F_T.F_\rho$ is larger than $3\%$. With this correction, Kierstead fit (unchanged) and Arp and McCarty fit (modified and extrapolated) reconnect much better.

We should mention here that $C_v$ cannot be derived from the state equation. In the zero-density region, $C_v$ is known exactly (perfect gas) and in the critical region, it has been fitted by Moldover\cite{Moldover1969}. In between, we resorted to exact thermodynamics relation to bridge to either one of these two regions.

\subsection{The transport properties : viscosity and thermal conductivity}

A fit of $^4$He viscosity in the range $4-20\ K$ and $0-10\ MPa$ has been proposed by Steward and Wallace\cite{Steward1971}. In our range of parameters, the fit is an interpolation of isothermal measurements at 4, 5, 6 and $10\ K$ conducted by these authors. Along the critical isochore, comparison with viscosity data of Kogan \textit{et al.}\cite{MeyerPrivate} and Agosta \textit{et al.}\cite{Agosta1987} shows $+7\%$ deviation at $5.2\ K$ and $-7\%$ at $7\ K$. However Steward and Wallace measurements at 4, 5 and $6\ K$ are in a few percent agreement with the literature, including the 2 references mentioned above, but their data at $10\,K$ differ significantly from the literature. It appeared that  this $10\ K$ isothermal entails a strong bias on Steward and Wallace fit down to the lower temperatures : this is consistent with a concern regarding a contamination of helium, due to a defective purifier\cite{ArpPrivate}. Consequently, we derived a new interpolation between the 4, 5 and $6\ K$ isothermals above $70\ kg/m^3$ and with additional data along the critical isochore\cite{MeyerPrivate}, in the zero-density limit (ab-initio calculation of ref.~\cite{Aziz1995}), and on the vapour-liquid curve\cite{Donnelly1998}. Concerning viscosity, there is a clear need of new measurements in the range $6K-10K$ and above $\rho_c$. We have no data in this range but this lack of information makes doubtful the interest in publishing our fit.
\\
The thermal conductivy has been estimated from a specially designed new fit through the data of Acton and Kellner\cite{Acton1977,Acton1981}. We re-computed the density data of these papers, which had been estimated with the 1973 fit of McCarty\cite{McCarty1973}, even in the critical region. Our new fit agrees within $\pm2\%$ with the published\cite{Acton1977,Acton1981} and unpublished\cite{MeyerPrivate} data of Acton and Kellner. Whenever it was possible, our convection measurements have been conducted at the same mean temperatures as the one employed by Acton and Kellner, in order to minimize interpolation errors.

\section{EXPERIMENTAL RESULTS AND ANALYSIS}

The remaining of this paper presents the heat transfer measurements and their consequences. First we consider the $Nu(Ra)$ relation. Our high accuracy on $Nu$ gives access to two new effects : the bimodality of $Nu$ and the side-wall conductivity influence on heat transfer. We then turn to the $Nu(Pr)$ dependence.

\begin{figure}
\centerline{\includegraphics[width=5.5in]{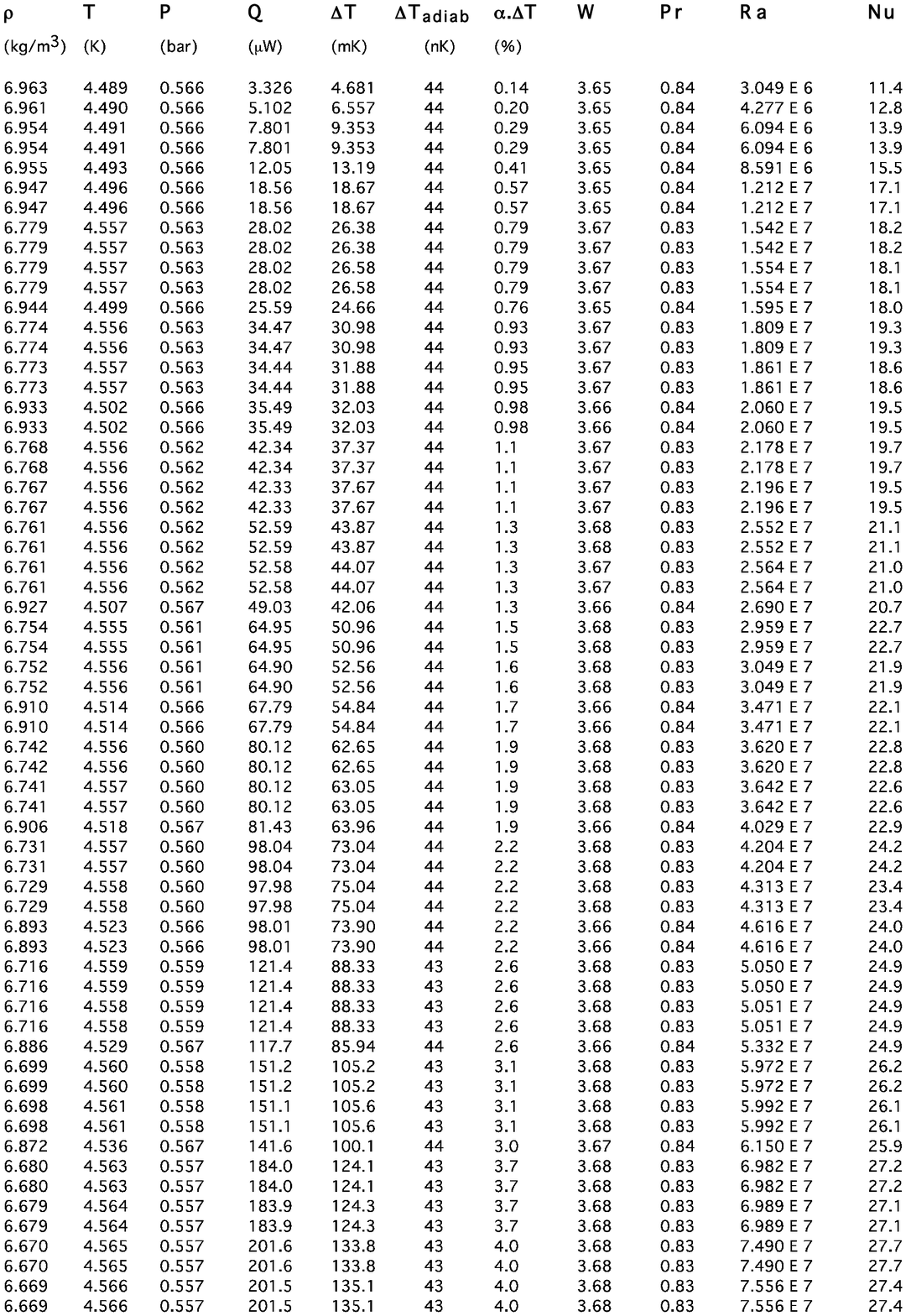}}
\label{fig:d1}
\end{figure}

\begin{figure}
\centerline{\includegraphics[width=5.5in]{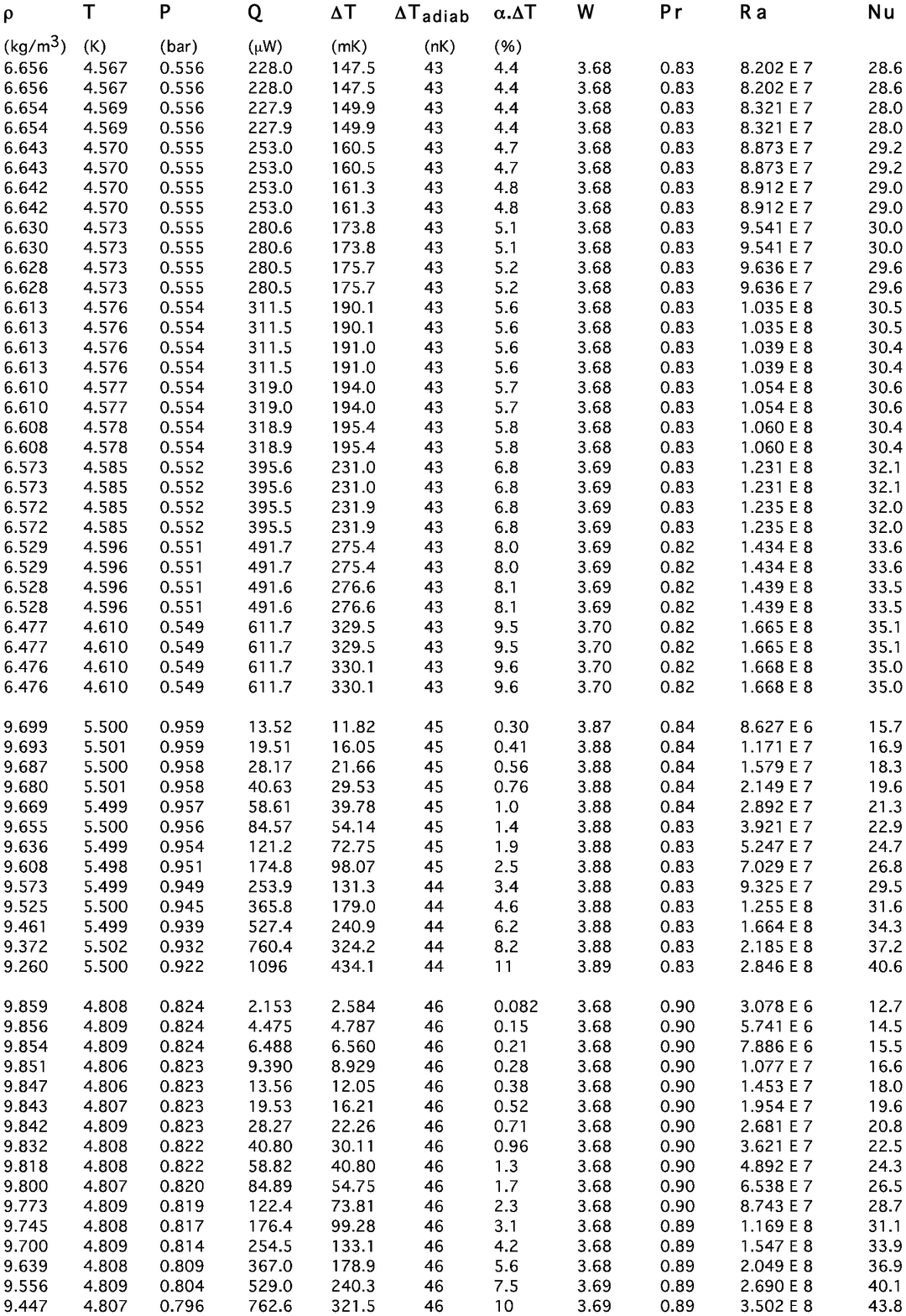}}
\label{fig:d2}
\end{figure}

\begin{figure}
\centerline{\includegraphics[width=5.5in]{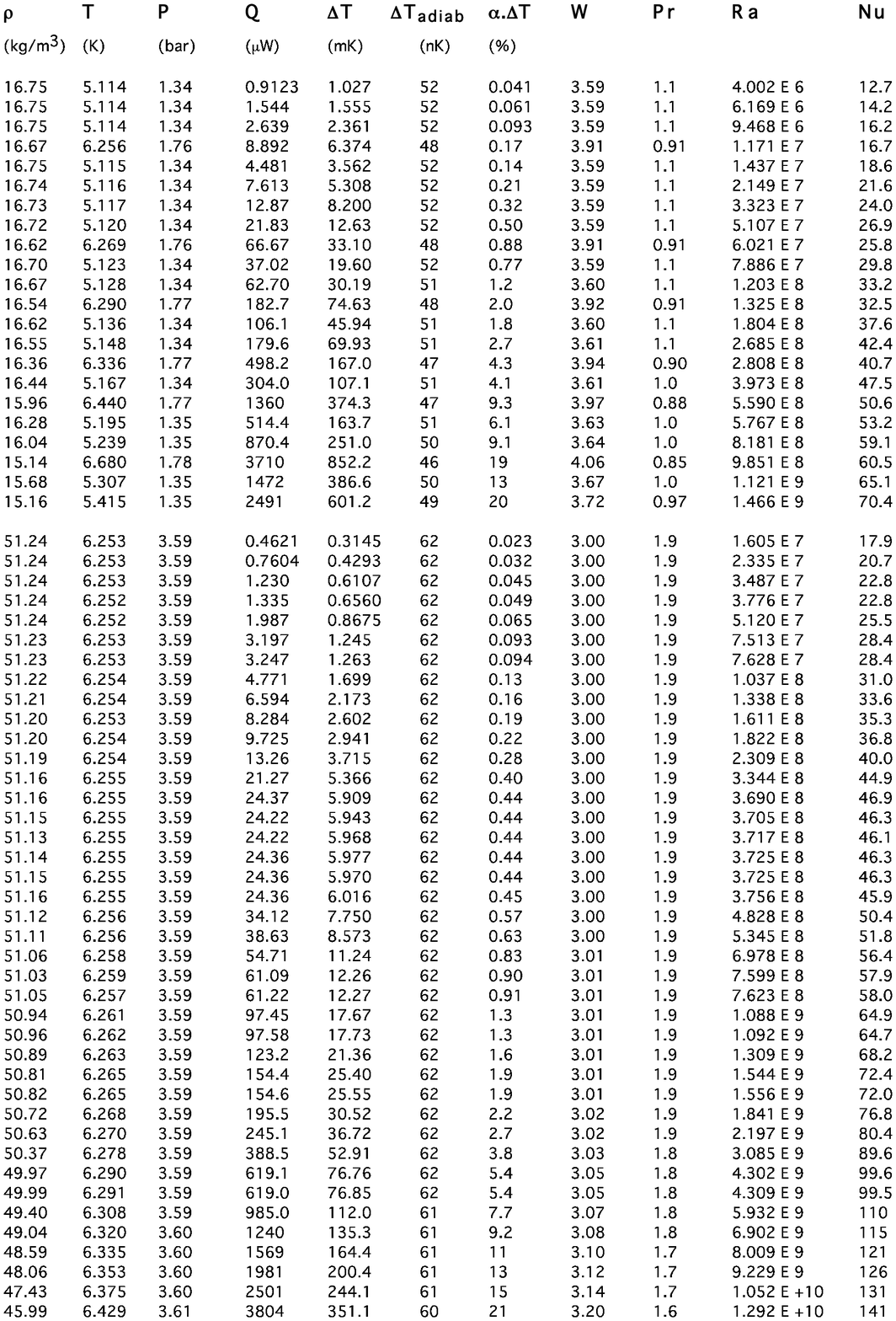}}
\label{fig:d3}
\end{figure}

\begin{figure}
\centerline{\includegraphics[width=5.5in]{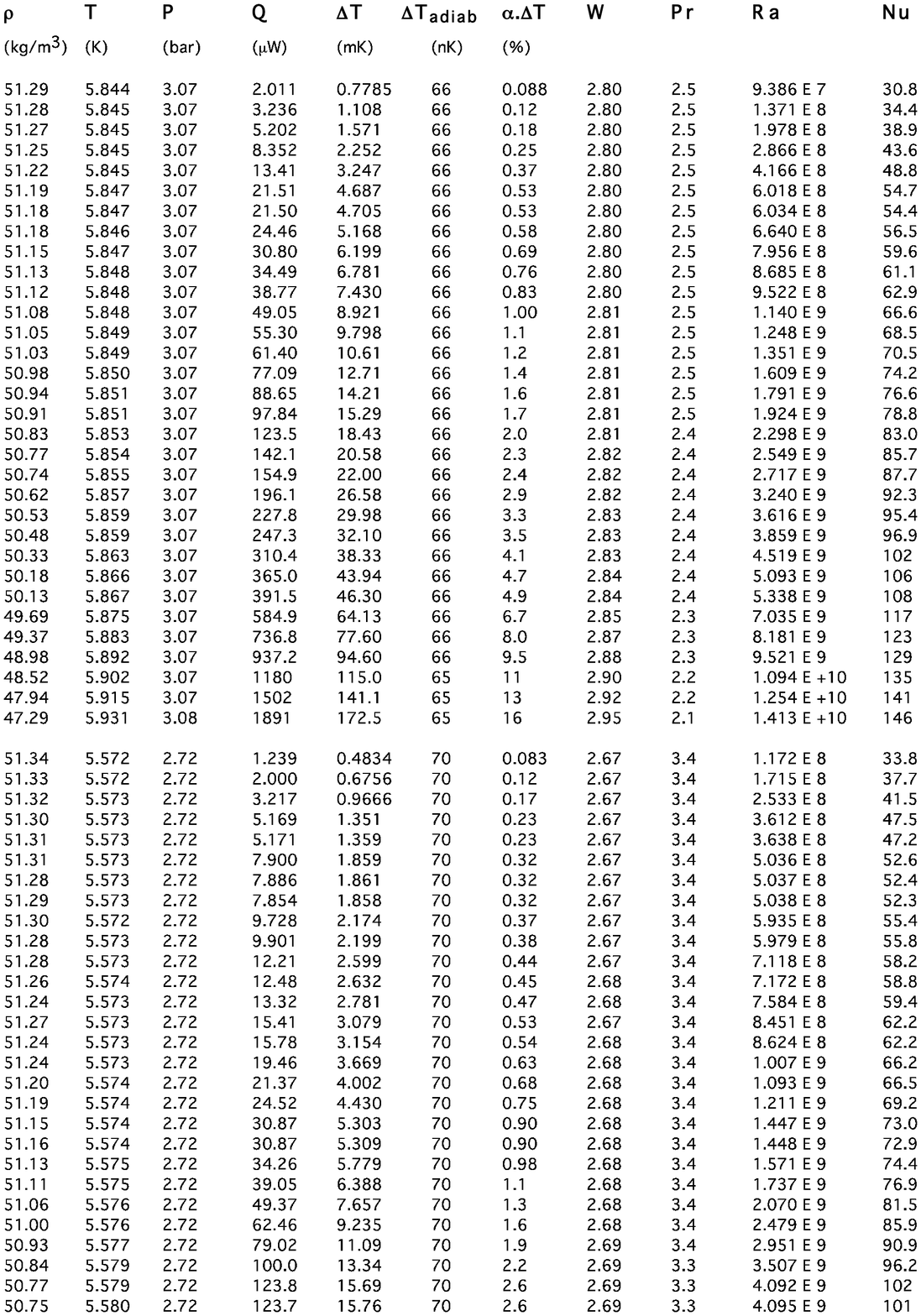}}
\label{fig:d4}
\end{figure}

\begin{figure}
\centerline{\includegraphics[width=5.5in]{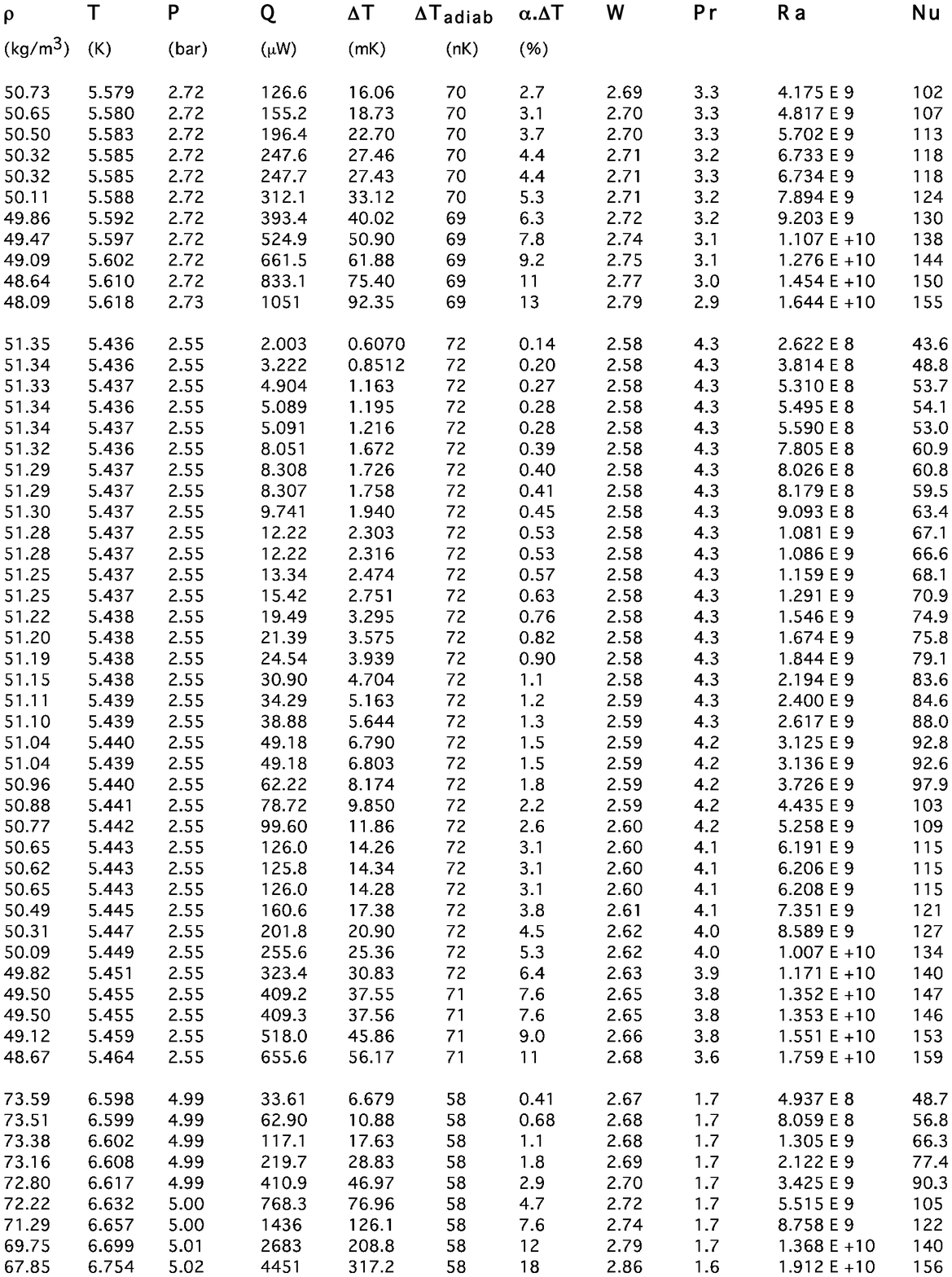}}
\label{fig:d5}
\end{figure}

\begin{figure}
\centerline{\includegraphics[width=5.5in]{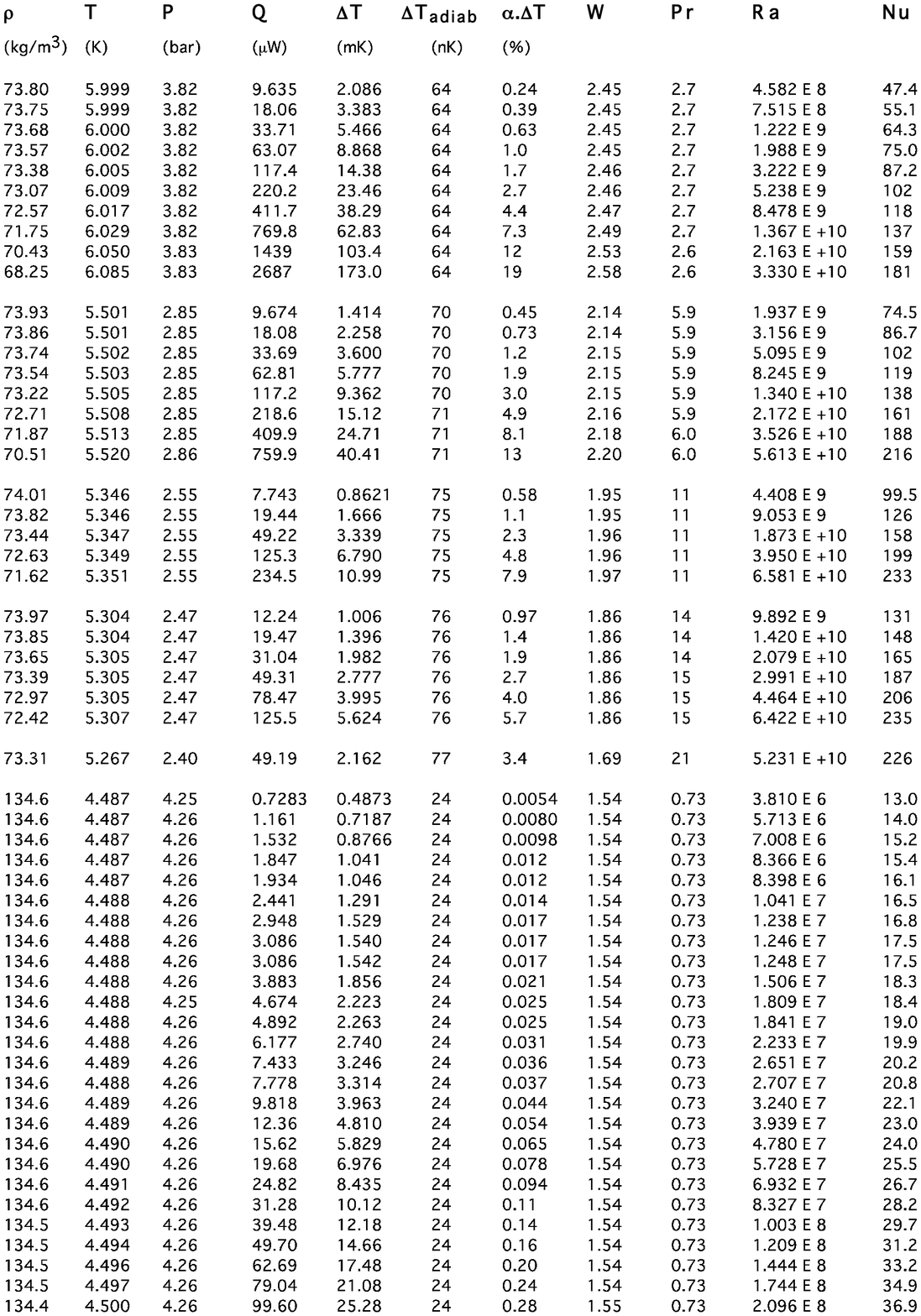}}
\label{fig:d6}
\end{figure}

\begin{figure}
\centerline{\includegraphics[width=5.5in]{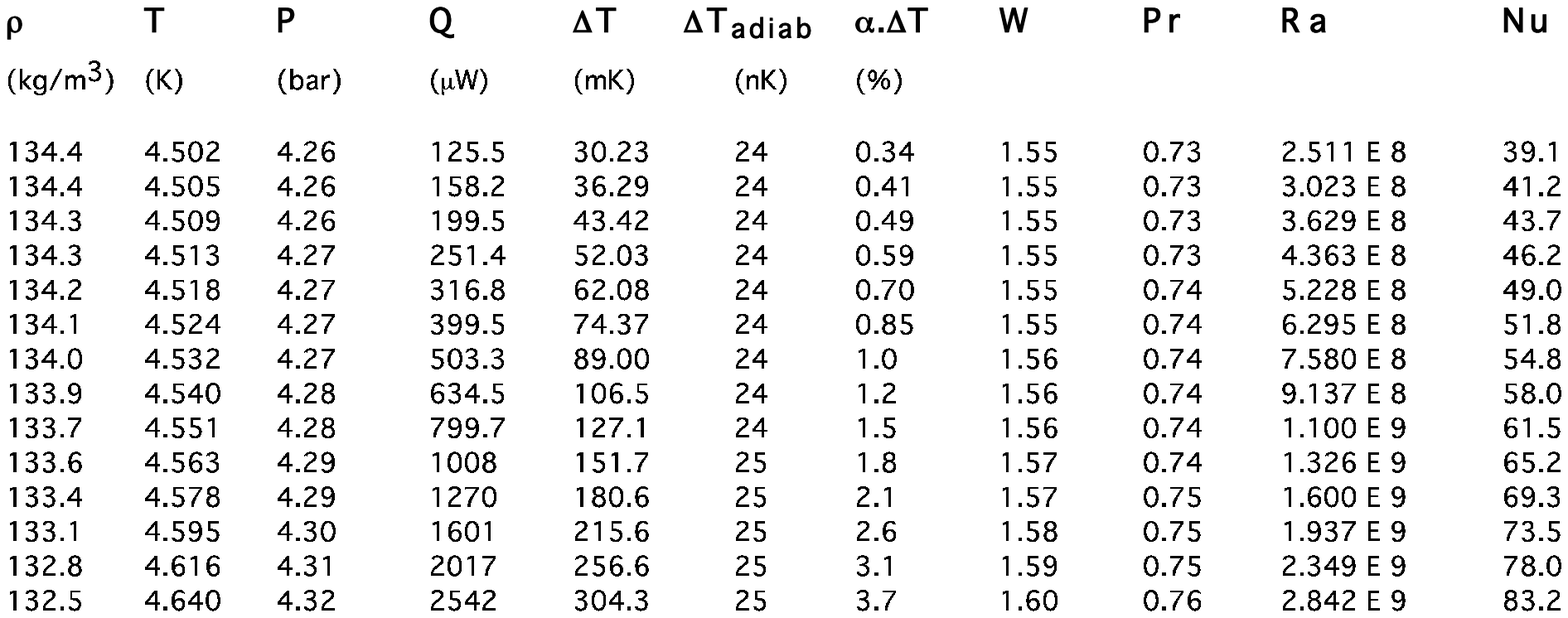}}
\label{fig:d7}
\end{figure}

\subsection{Bimodality}

\begin{figure}
\centerline{\includegraphics[width=3.5in,angle=90]{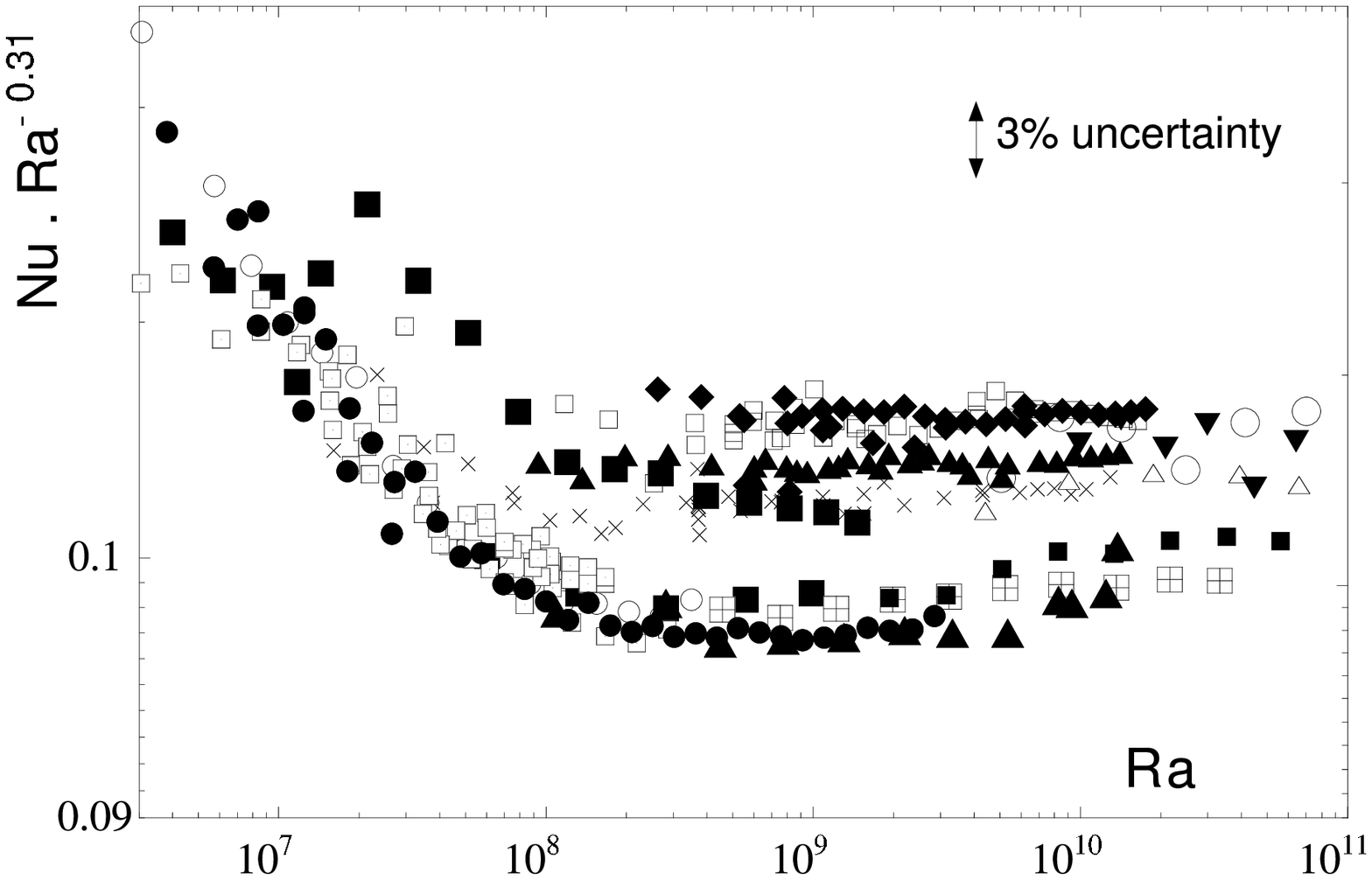}}
\caption{
$NuRa^{-0.31}$ vs. $Ra$ in the mini-cell.
\mbox{\Large $\bullet$} $Pr\simeq 0.73-0.76$
\mbox{\scriptsize $\boxdot$}  $Pr\simeq 0.82-0.84$,
\mbox{\Large $\circ$} $Pr\simeq 0.89-0.90$,
\mbox{\large $\blacksquare$}  $Pr\simeq 0.85-1.1$
\mbox{\footnotesize $\times$}  $Pr\simeq 1.6-1.9$,
\mbox{\scriptsize $\blacktriangle$}  $Pr\simeq 2.1-2.5$,
\mbox{\large $\boxplus$} $Pr\simeq 2.6-2.7$,
\mbox{\scriptsize $\square$}  $Pr\simeq 2.9-3.4$,
\mbox{\footnotesize $\blacklozenge$} $Pr\simeq 3.6-4.3$,
\mbox{\tiny $\blacksquare$}  $Pr\simeq 5.9-6.0$,
\mbox{\footnotesize $\triangle$}  $Pr\simeq 11$,
\mbox{\large $\blacktriangledown$} $Pr\simeq 14-15$,
For comparison, large cell\cite{RochePRE2001} data are plotted with the symbols :
\mbox{\Large $\blacktriangle$} $Pr\simeq 0.7$,
\mbox{\huge $\circ$}  $Pr\simeq 1.1$.
}
\label{fig:bimodality}
\end{figure}

On figure~\ref{fig:bimodality} we show $Nu$ as a function of $Ra$. In order to display all data with an increased vertical resolution, $Nu$ is arbitrarily re-normalized by $Ra^{0.31}$. For comparison, we display the data at $Pr= 0.7$ and $1.1$  from ref.~\cite{RochePRE2001}. The change in slope at $Ra\simeq 10^8-2.10^8$ corresponds to the soft-hard turbulent transition in $1/2$ aspect ratio cell\cite{WuTHESE}. In the soft regime, the exponent of $Nu$ vs. $Ra$ is close to 0.25 (interpolation over only 1.5 decade), while in the hard turbulence regime the exponent is close to 0.31, in between the $2/7$ and $1/3$ predictions of traditional theories\cite{Siggia1994}. It is interesting to note that the exponent averaged over these two regimes is close to $2/7$. Another possible interpretation of the exponent change would be to reject the soft-hard transition picture and rather see a continuous variation of the exponent resulting from a linear combinaison of two power laws, although the abruptness of the exponent change doesn't fully fit in this picture.

For $2.10^7 < Ra < 2.10^{10}$, points can be gathered into two subsets which differ by roughly $5-7\%$ in $Nu$. Such a data bimodality cannot be taken into account by the uncertainties, which are twice smaller than this $Nu$ gap, nor by a $Pr$ dependence. Switches from one set of data to the other occur varying $\Delta T$ under quasi-constant mean temperature and density conditions (see for example $Pr\simeq 0.95$) : this definitely rules out that the bimodality would come from an improper helium property estimation in the T-$\rho$ plane.%Such a bimodality can also be seen in the large cell data from ref.~\cite{RochePRE2001}, as shown on figure~\ref{fig:ultime}.
 A numerical simulation conducted for the same cell geometry\cite{Verzicco2003} recently found that two types of large scale flow can fit the cell. This mechanism of bimodality is consistent with the invariance of the $Nu$ gap (in log scale) observed in our data. 

We cannot decide if the bimodality reveals spurious effect of the boundary conditions or a macroscopic degree of freedom of the flow with a slow dynamics. In the first hypothesis, each mode could be stabilized by the thermal inertia of cell boundaries (for instance, an ascending wind warms up the nearby side-wall which -in turns- enhance an ascending convection) and the switching from one mode to the other should be hysteretic. We are not able to precise more what is the anchoring mechanism. The second hypothesis has drastic consequences since the very slow dynamics (at least hundreds of turn-over times) ruins -for practical reasons- the present definition of $Nu$ : indeed a clean averaging procedure would request a time duration incompatible with a laboratory experiment.

In this paragraph, we present a practical difference between the cryogenic and room temperature experiments. These latter ones are illustrated quantitatively with water as the working fluid. The characteristic time of convection is :
\begin{equation}
t=h^2/\kappa=\sqrt{\frac{Ra.Pr.h}{\alpha .\Delta T.g}}
\label{t }
\end{equation}

At a given $Ra$ and $Pr$, the caracteristic time $t_{He}$ and $t_{water}$ in helium and water obeys to :
\begin{equation}
\frac{t_{water}}{t_{He}} = \sqrt{\frac{h_{water}}{h_{He}}.\frac{(\alpha .\Delta T)_{He}}{(\alpha .\Delta T)_{water}}}
\label{HeWater }
\end{equation}

with obvious notations. In helium and water cells, the smallest convection times are obtained at the highest $\alpha .\Delta T$ which -given Boussinesq conditions- are typically $(\alpha .\Delta T)_{He}=0.2$ and $(\alpha .\Delta T)_{water}=0.02$. The $Ra$ explored in this $h_{He}=2$~cm experiment are achievable with water for $h_{water}=20$~cm typically, which gives ${t_{water}}/{t_{He}}=10$. We can state more generally that the times scales in cryogenic helium are typically 10 times smaller than in water for given $Ra$ and $Pr$. More than  330 data points are plotted on figure~\ref{fig:bimodality}, each $Nu$ has been averaged over typically 1 hour and after a relaxation time of a few hours ($t_{He}$ is a few tens of seconds). Obtaining so many data would have required about one year and half of continuous operation in water. If the bimodality results from a macroscopic degree of freedom, it is likely that the observation of two modes was possible in our experiment  thanks to the small times scales involved.

Let us now present a consequence of this bimodality for room temperature convection cells, assuming that bimodality results from the slow-dynamics degree of freedom. In heat transfer measurements or visualisation, the possibility to be in a slow transient mode-flipping should be considered. The resulting transient-regime uncertainty may be a delicate experimental issue for flow characterization studies, and a limiting factor for precise measurements, at least without new specific cell design.

\subsection{Side-Wall conductivity}

A second outcome of this experiment was the finding of a significant and unexpected influence of the side-wall conductivity on the apparent Nusselt number, as also discussed in references~\cite{AhlersWall2001,Verzicco2002,Niemela2003} : typical side-walls found in the literature are responsible for overestimation of $Nu$ up to more than $20\%$. We have conducted a specific study, both theoretical and experimental for thin wall cells\cite{RocheEPJB2001}.

\begin{figure}
\centerline{\includegraphics[height=2.5in]{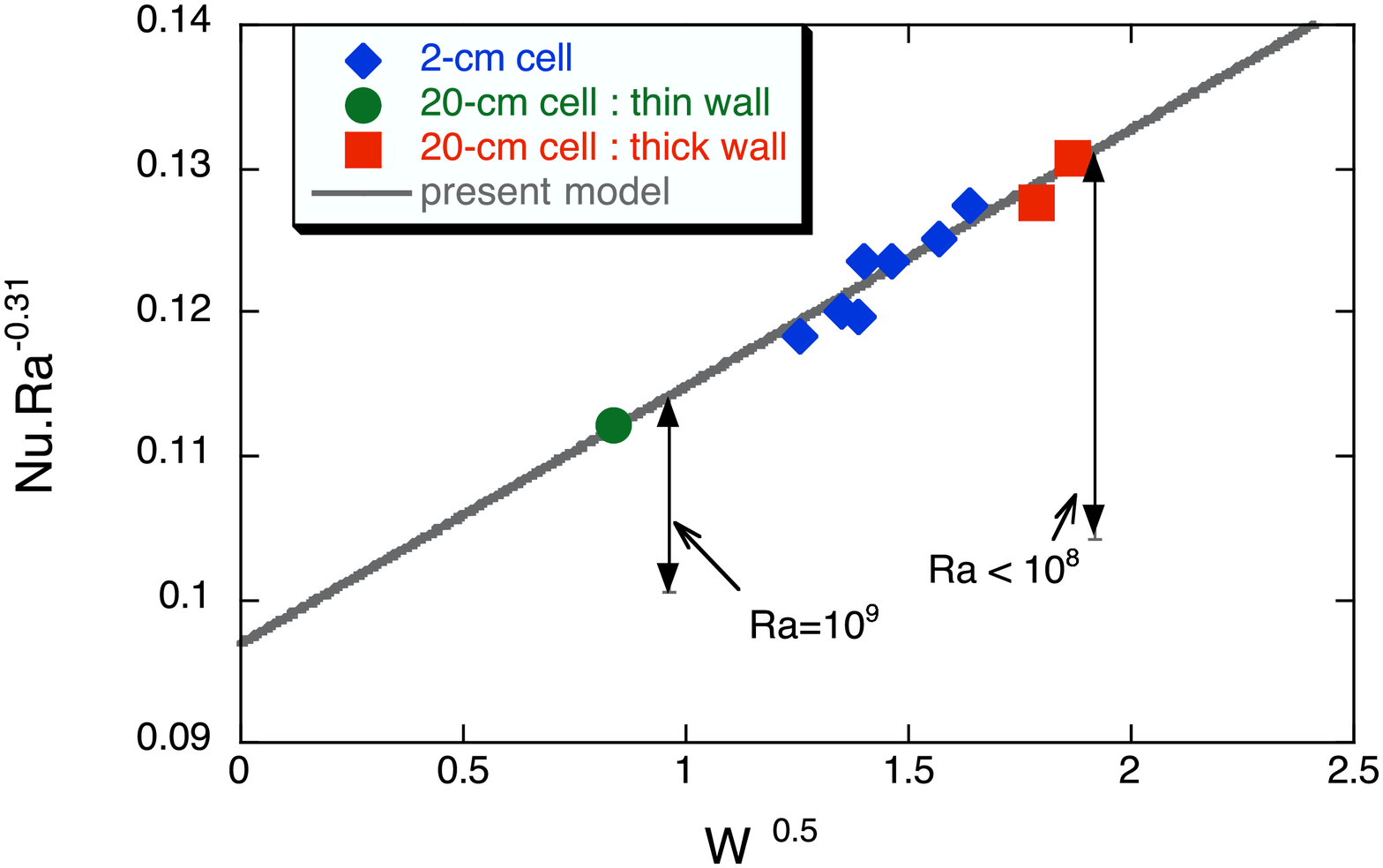}}
\caption{$Nu.Ra^{-0.31}$ vs. the square root of the side-wall number W. The symbols are experimental data obtained in the lower mode, for $10^9<Ra<5.10^9$ in $h=2\,cm$ and $20\,cm$ cells. The vertical bars represent the magnitude of the side-wall correction estimated from numerical simulations by Verzicco\cite{Verzicco2002}. The solid line is the present analytical model with one adjustable parameter.}
\label{fig:wall}
\end{figure}

As discussed in ref.~\cite{RocheEPJB2001}, the influence of thin walls can be characterized by a dimensionless number, called the side-wall number $W$. This number is the ratio between the conductance of the empty cell and that of the fluid at rest. Typical values of $W$ for reference experiments range from nearly 0 up to 3.5. Points on figure~\ref{fig:wall} gather our heat transfer measurements (restricted to the lower ``modality'').

The starting point are the same as reference~\cite{AhlersWall2001} : the wall is in contact with a fluid of nearly uniform temperature $T$ through a lateral boundary layer. In contrast with reference~\cite{AhlersWall2001}, we do not take the the thickness of the boundary layer as constant (model 2 of~[\cite{AhlersWall2001}])  nor proportional to $1/\sqrt{Re}$ (model 1 of~[\cite{AhlersWall2001}]) but proportional to the thermal boundary layer of the plates, that is $1/Nu$. It results in an exchange length between the wall and the bulk proportional to $1/\sqrt{Nu}$. Another difference with reference~\cite{AhlersWall2001} is that we consider this exchange length as extending the effective area of the plate, instead of considering the calculated flux as a thermal leak. This point of view is clearly confirmed by numerical simulation\cite{Verzicco2002}.  A $Nu$ correction based on the leak, rather than corrected exchange surface, gives a poorer fit on the experimental data (less of 2 decades of $Ra$ are well corrected).Our picture result in an asymptotic correction in $\sqrt{W}$ fully confirmed by our compilation~\ref{fig:wall}.

The closed analytical correction formula we derived has an adjustable parameter accounting the proportionality between the lateral boundary layer thickness and the plates' thermal ones.

We derived a closed-analytical correction formula for the Nusselt number with one adjustable parameter (continuous line of figure~\ref{fig:wall}). Calling $Nu_{mea}$ and $Nu_{cor}$ the measured and corrected Nusselt numbers :

\begin{equation}
Nu_{mea}=Nu_{cor}(1+f(W))
\label{W 1}
\end{equation}

with

\begin{equation}
f(W)=\frac{A^2}{\Gamma Nu_{cor}}(\sqrt{1+\frac{2W\Gamma Nu_{cor} }{A^2}}-1)
\label{W 2}
\end{equation}

and

\begin{equation}
A=0.8 
\label{W 3}
\end{equation}

The dependence of this correction with the Nusselt itself has been validated for Rayleigh numbers covering 4 decades (and for various values of $W$). On the figure, the vertical arrows represent the correction magnitude numerically estimated by Verzicco\cite{Verzicco2002}. 

The present analysis has numerous consequences. First it explains some surprising results and discrepancies between published results, as detailed in ref.~\cite{RocheEPJB2001}. Second, the correction on $Nu$ being $Ra$ dependent, it changes the $Nu(Ra)$ apparent exponent, at least for $Ra<10^{10}$. The examination of several published results indicated that the measured exponent -often close to $2/7$- are significantly underestimated. Values larger than 0.3 are obtained after correction (some experiments claiming a $2/7$ exponent are not subject to side-wall correction\cite{RocheTHESE}). Thirdly, it should be noted that the side-wall effect mimics a $Ra$ dependent exponent, which could be falsely interpreted as an indication of non-power law behaviour of convection. Finally, the wall-fluid interaction is likely to introduce a new length scale in the convection problem if the wall thickness is non-uniform (flanges, large o-ring,...). Such artefact could also cause apparent non-power law behaviour.

On the other hand, we can consider as very good news that such an
important effect can be corrected. It could modify the flow itself in such
a way that no comparison would be possible between the unperturbed and the
perturbed case. As shown by Verzicco\cite{Verzicco2002}, it is indeed the case when the
conductance of the wall is too high. Our experimental study, presented on
fig.~\ref{fig:wall}, shows that the wall number well correlates the various data which
shows the pertinence of the correction.

All the data presented in this paper are side-wall corrected. Note that the magnitude of the bimodality presented above is not affected by this correction.

\subsection{Prandtl number dependence}

Our experiment has been designed to study the influence of Prandtl number near the diffusivity cross-over $Pr \simeq 1$. We find a very small -if any- Prandtl number dependence over 1.5 decade\cite{RocheEPL2002}. This dependence corresponds to an exponent smaller (in absolute value) than 0.03 in a power law picture. This result is compatible with ref.~\cite{Ahlers2001} but not with ref.~\cite{Ashkenazi}. Also, the $2/7$ theories\cite{Castaing1989,Shraiman} predict an exponent $-1/7\simeq -0.14$ which is incompatible with our result.

\begin{figure}
\centerline{\includegraphics[height=3.5in]{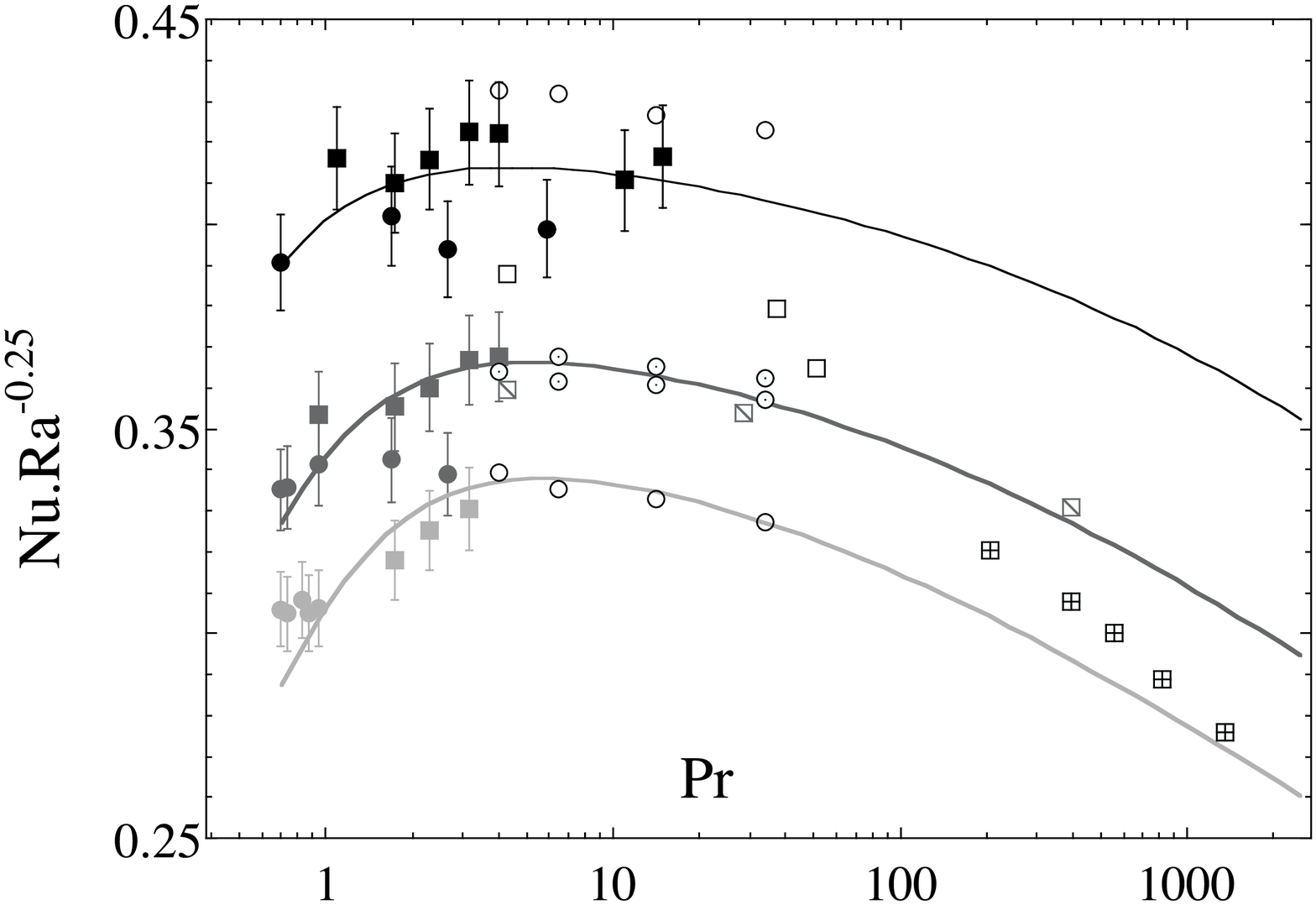}}
\caption{
$Nu.Ra^{-1/4}$ versus $Pr$ with -from bottom to top- $Ra=10^{8.25}$ (light gray), $Ra=10^9$ (dark gray) and $Ra=10^{10}$ (black). The value $Ra=10^{8.25}$ corresponds to the beginning of the hard turbulence regime for aspect ratio $\Gamma=0.5$. \mbox{\large $\bullet$} present data in the lower mode, \mbox{\tiny $\blacksquare$}  present data in the upper mode,  $\circ$ \mbox{\tiny $\odot$} data from ref.~\cite{Ahlers2001}, $\Gamma=0.5$ and $1$ (  \mbox{\tiny $\odot$} corresponds to $Ra=10^9$),  \mbox{\scriptsize $\square$} \mbox{\scriptsize $\boxminus$}  \mbox{\scriptsize $\boxplus$}  data from ref.~\cite{Xia2002} , $\Gamma=1$ (\mbox{\scriptsize $\boxplus$} : $Ra=10^{8.25}$, \mbox{\scriptsize $\boxminus$} : $Ra=10^9$). The lines are from Grossmann and Lohse\cite{Grossmann2001} model with adjustable parameters tuned by these authors on the $\circ$ \mbox{\tiny $\odot$}  data restricted to aspect ratio $\Gamma=1$.
}
\label{fig:lohse}
\end{figure}

A comparison with the prediction of Grossmann and Lohse\cite{Grossmann2001} (G.L.) can be performed using their 5 fitting parameters adjusted to fit the data of ref.~\cite{Ahlers2001}. It should be emphasized that the $Ra$ and $Pr$ overlap between the data of ref.~\cite{Ahlers2001} and ours is limited as shown on figure~\ref{fig:lohse}. Consequently we are testing a prediction of the G.L. theory : an extrapolation on the lower $Pr$ side. We find a good agreement since this prediction falls within the error bar of nearly all of data (figure~\ref{fig:lohse}). We should mention here another recent test~\cite{Xia2002} of G.L. theory on the higher $Pr$ side. These data are also in good agreement with the prediction as shown on figure~\ref{fig:lohse}. These three set of data are for two different aspect ratio ($\Gamma=0.5$ and $1$), but once the wall effect corrected, the influence of the aspect ratio seems weak\cite{AhlersWall2001}. In particular, all the set of data from the different groups suggest that the transition from the low Prandtl regime to the high Prandtl one occurs in the neighbourhood of $Pr\simeq 1$ and not $0.1$ as proposed by Kraichnan\cite{Siggia1994,Kraichnan1962}. 

\section{CONCLUDING REMARKS}

We have measurement $Nu(Ra)$ and $Nu(Pr)$ dependences which are both incompatible with the 2/7 and 1/3 theories\cite{Siggia1994}, at least under their present form. The Grossmann and Lohse theory\cite{Grossmann2001} can account for our data but the discriminating testing of the 5 fittings parameters have to be made on a larger range of $Ra$ and $Pr$ numbers.

The bimodality effect indicates that the mean flow confinement has a significant influence (up to few percents) on the precise $Nu(Ra)$ dependence and this influence should hold for all aspect ratio of order 1. Since confinement effects (multi-modality, poor spatial homogeneity on boundary layers, ...) are not considered by present theories, their predicting power is indeed limited in precision. This darkens the perspective that very precise $Nu(Ra)$ measurements can discriminate between competing theories. This conclusion is reinforced by the boundary conditions influence on the global heat transfer, such as side-wall conductivity \cite{RocheEPJB2001,AhlersWall2001,Verzicco2002} and hole-burning effects in plates\cite{CastaingETC82000,ChaumatETC,Verzicco_plate}. This underlines the importance of alternative approaches to probe the heat transfer mechanism and (in)validate theories. It also calls for a new generation of cell design with a specific attention dedicated to the mean flow and thermal boundary conditions. For example, the influence of the large scale flow on the heat transfer suggests that large aspect ratio cell could be required to observe true power law scalings.

\section*{ACKNOWLEDGMENTS}
Ackowledgments are due to J. Niemela, J. Sommeria and R. Verzicco for fruitful discussions, and to V. Arp and H. Meyer for help in retrieving literature on helium properties.

\end{document}